\documentclass[%
 reprint,
 amsmath,amssymb,
 aps,
]{revtex4-2}
\usepackage{xcolor}
\usepackage{overpic}
\usepackage{graphicx}
\usepackage{dcolumn}
\usepackage{bm}
\usepackage{tikz}
\usetikzlibrary{tikzmark,calc}
\usepackage{float}

\usepackage{ragged2e}
\begin{document}

\preprint{APS/123-QED}

\title{Shear-Induced Phase Behavior and Topological Defects in Two-Dimensional Crystals}

\author{Federico Ghimenti}%
\affiliation{Laboratoire de Physique de l'Ecole Normale Sup\'erieure, ENS, Universit\'e PSL, CNRS, Sorbonne Universit\'e, Universit\'e Paris Cit\'e, F-75005 Paris, France}
\affiliation{Laboratoire Matière et Systèmes Complexes, UMR No. 7057, CNRS, Université Paris Cité, 10 rue Alice Domon et Léonie Duquet, 75013 Paris, France}

\author{Misaki Ozawa}%
\affiliation{Laboratoire de Physique de l'Ecole Normale Sup\'erieure, ENS, Universit\'e PSL, CNRS, Sorbonne Universit\'e, Universit\'e Paris Cit\'e, F-75005 Paris, France}
\affiliation{Univ. Grenoble Alpes, CNRS, LIPhy, 38000 Grenoble, France}

\author{Giulio Biroli}%
\affiliation{Laboratoire de Physique de l'Ecole Normale Sup\'erieure, ENS, Universit\'e PSL, CNRS, Sorbonne Universit\'e, Universit\'e Paris Cit\'e, F-75005 Paris, France}

\author{Gilles Tarjus}%
\affiliation{LPTMC, CNRS-UMR 7600, Sorbonne Universit\'e, 4 Place Jussieu, F-75005 Paris, France}

\date{\today}

\begin{abstract}
We investigate through numerical simulations how a two-dimensional crystal yields and flows under an applied shear. We focus over a range that allows 
us to both address the response in the limit of an infinitesimal shear rate and describe the phase behavior of the system at a finite shear rate.  In doing so, we carefully 
discuss the role of the topological defects and of the finite-size effects. We map out the whole phase diagram of the flowing steady state in the plane formed by 
temperature and shear rate. Shear-induced melting of the two-dimensional crystal is found to proceed in two steps: first, the solid  loses long-range bond-orientational order 
and flows, even for an infinitesimal shear rate (in the thermodynamic limit). The resulting flowing hexatic phase then melts to a flowing, rather isotropic, liquid at a 
finite shear rate that depends on temperature. Finally, at a high shear rate, a third regime corresponding to a strongly anisotropic string-like flowing phase appears.
\end{abstract}

\maketitle


\section{Introduction}

How do crystals flow under an applied shear? This question can be viewed from two different perspectives. Alternatively, one may envisage the onset 
of flow as an instance of a {\it yielding transition} between an elastically responding rigid solid and a plastically flowing phase~\cite{sethna2017deformation}. This pertains to a broad field 
of research within mechanics,  soft-condensed matter and statistical physics which involves a very wide range of materials from granular media, foams, and a 
whole variety of so-called yield-stress fluids to all kinds of harder solids such as glasses and to crystalline materials~\cite{bonn2017yield,alava2006statistical}. One is then concerned with the mechanisms 
inducing plasticity, the properties of the flow, the existence and the value of the yield stress, the nature of the yielding transition, and all means to control the 
way the solids yield without breaking too soon. One may also consider the phenomenon in a more specific way as a {\it shear-induced melting transition} associated 
with some symmetry restoration and enquire how this transition proceeds and differs (or not) from the melting of the quiescent crystal in equilibrium~\cite{chaikin1995principles}. 

Plasticity in crystals is known to be due to the presence of defects in the structure, above all topological defects in the form of dislocations. In many real systems 
they are present in a rather large quantity and, having been trapped in the solid during its preparation, they are out of equilibrium. Here instead we are interested in starting with  
{\it perfect equilibrium crystals},  which, as a result, only contain thermal topological defects compatible with the fixed nonzero temperature. We focus on the steady 
state reached by imposing a constant shear (strain) rate and do not address transient effects that may give a different angle on the yielding transition. Furthermore, 
we consider a two-dimensional crystal, as for instance experimentally studied in colloidal suspensions~\cite{stancik2004dynamic,gasser2010melting}, hexagonal columnar liquid crystals~\cite{ramos2004shear}, complex plasmas~\cite{ivlev2012complex}, and for which more analytical work is possible in the context of the KTNHY theory of 
melting~\cite{kosterlitz1972long,kosterlitz2018ordering,halperin1978theory,nelson1979dislocation,young1979melting}. In two dimensions the crystal  has only quasi-long-range translational (crystalline) order but long-range bond-orientational order.  (Note that here and below we use for convenience the terminology ``crystal'' even in two dimensions where there is no long-range translational order; this is an abuse of language but should not lead to any confusion.) Melting in 
equilibrium may take place through two distinct transitions that are associated with the unbinding of bound topological defects and are separated by an 
intermediate ``hexatic'' phase. The crystal-to-hexatic transition corresponds to the appearance of free dislocations, and the resulting hexatic phase only has 
quasi-long-range bond-orientational order. The hexatic-to-liquid transition corresponds to the unbinding of the dislocations into free disclinations which therefore also 
break the quasi-long-range order and fully restore translational and bond-orientational invariance.

Our goal is to investigate how a two-dimensional crystal yields and flows under an applied shear over a range of rates that allows us to both address the 
response in the limit of an infinitesimal shear rate and describe the phase behavior of the system at finite rate. It has been theoretically 
established~\cite{sausset2010solids,nath2018existence,reddy2020nucleation} that even a perfect crystal flows for an infinitesimal shear so that the notion 
of yield stress is only a time-dependent property which should vanish for a large, yet finite, observation time (even in the thermodynamic limit). A viscosity can then 
be defined but it diverges in a singular manner for a vanishing shear rate. We give numerical evidence for these predictions and discuss the mechanism 
by which this takes place in two-dimensional crystals. For larger shear rates we provide a description of the shear-induced melting and of the properties of the 
phases that are observed in a steady state.

\section{Model, method, and phase diagram}
\label{sec:model}

We numerically study a model of dense monodisperse colloidal crystals under simple shear in two dimensions. We consider the situation where hydrodynamic 
interactions and inertial effects can be neglected and we perform a Brownian (overdamped Langevin) dynamics for the position $\mathbf{r}_i=(x_i, y_i)$ of each particle 
under a constant and uniform applied strain rate $\dot\gamma$~\cite{ikeda2012unified}:
\begin{equation}
    \zeta \frac{\mathrm{d} \mathbf{r}_i}{\mathrm{d} t} = -\sum_{j \neq i} \frac{\partial v(\mathbf{r}_i - \mathbf{r}_j)}{\partial \mathbf{r}_i} + \dot\gamma \mathbf{e}_x y_i + \mathbf{f}_i,
    \label{eq:langevin}
\end{equation}
with $v(\mathbf{r})=\frac{\epsilon}{2} (1 - |{\bf r}|/d )^2 \theta(d-|{\bf r}|)$ a purely repulsive soft potential, where $d$ is the particle diameter and $\theta(x)$ is the 
step function. The thermal bath is described through the stochastic force $\mathbf{f}_i=(f_{x, i},f_{y, i})$, which is a Gaussian white noise with zero mean and correlations given by 
$\langle  f_{\alpha, i}(t)f_{\beta, j}(t')\rangle = 2k_B T\zeta\delta(t-t')\delta_{ij}\delta_{\alpha \beta}$, where $\langle \cdots \rangle$ is a statistical average, $T$ is the 
temperature of the bath, $k_B$ is the Boltzmann constant, and $\alpha, \beta=x, y$. We measure lengths in units of the diameter $d$, times in units of 
$\tau_0=\zeta d^2/\epsilon$, and temperature in units of $\epsilon/k_B$.

We study $N$ harmonic soft disks in a rectangular box with  area $A=L_x L_y$, where $L_x$ is the box length along the $x$-direction and $L_y=\frac{\sqrt{3}}{2}L_x$ 
is the length along the $y$-direction. The ratio is chosen to accommodate the perfect hexagonal structure. The packing fraction $\phi$ of the system is set to 
$\phi =(N/A)\pi d^2/4= 1.0$, for which the system has been shown to have a first-order hexatic-to-liquid transition~\cite{engel2013hard} at $T_{m,{\rm hex}} \simeq 0.0062\pm 0.0002$ in 
thermal equilibrium without applied deformation ($\dot\gamma=0$)~\cite{zu2016density}. Although the full equilibrium phase diagram of the model for $\dot\gamma=0$ 
is not available, we note that the hexatic phase in soft-core potential models always appears in a narrow range of temperature (or density, but for power-law potentials 
the latter can easily be converted to temperature) which is a few percents of the transition temperature of the hexatic phase to the 
liquid~\cite{zu2016density,kapfer2015two}: we therefore estimate the melting temperature of the solid to the hexatic phase to be 
$T_{m,{\rm sol}} \gtrsim 0.0055$.

To implement the uniform simple shear, Lees-Edwards periodic boundary conditions are applied~\cite{allen2017computer}, and the equations of motion are 
integrated through the Euler scheme. We measure the shear stress component of the system, $\sigma = \sigma_{xy}$, by using the Irving-Kirkwood formula~\cite{allen2017computer}: see Appendix~\ref{app:shear-stress}. In the initial condition, particles are arranged in a hexagonal close-packed 
structure, which is then subjected to an applied shear at the chosen temperature based on Eq.~(\ref{eq:langevin}). All the quantities presented in this paper are measured {\it in the steady state} (after a long enough simulation time), except otherwise 
stated. We investigate a wide range of shear rate $\dot \gamma$ and temperature $T$, which covers most of the relevant physics of two-dimensional ($2d$) 
crystal flows and we study $N=900$, $3600$, $14400$, and $57600$ to check the finite-size effects.

Note that we consider a Brownian (overdamped Langevin) dynamics which is appropriate for colloidal suspensions and is different from the previous simulation studies of sheared two-dimensional crystals that used a nonequilibrium molecular dynamics algorithm (SLLOD)~\cite{weider1993shear,delhommelle2004simulations}. In the 
latter case there is an issue concerning the way the system is thermostated (kinetic or a configurational thermostat), which may influence some of the 
results~\cite{delhommelle2004simulations}. This specific problem is absent in our Brownian dynamics simulations where temperature is introduced through a white 
noise. For completeness we have also carried out SLLOD dynamics simulations: the results are discussed in Appendix~\ref{app:SLLOD}.

\begin{figure*}
\centering
\includegraphics[width=1.8\columnwidth]{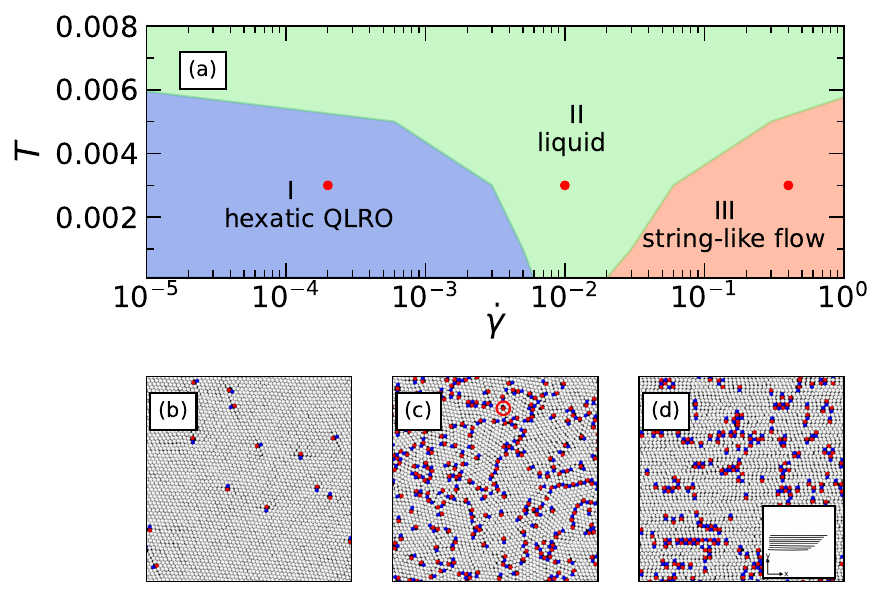}
\caption{(a): Phase diagram of a sheared two-dimensional crystal in its flowing steady state in the plane of the shear rate $\dot \gamma$ and the temperature $T$. 
Red dots indicate the phase points corresponding to the snapshots displayed in panels (b-d). 
In Regime I,  we observe a plastic flow with nucleated free dislocations and hexatic quasi-long-range order (QLRO). A representative snapshot is shown in (b) 
for $T=0.003$ and $\dot \gamma = 2\times10^{-4}$. Blue, white, and red particles have 5, 6, and 7 neighbors, respectively, and a pair of red and blue particles form 
a dislocation. 
In Regime II, the dislocations are unbound and free disclinations, shown as isolated red and blue particles, are nucleated. Concomitantly, bond-orientational order has a 
short-ranged, exponential, spatial decay and the system is in a flowing liquid phase. A representative snapshot is given in (c) for $T=0.003$ and 
$\dot \gamma = 1\times10^{-2}$. One can see an isolated disclination with 7 neighbors, as indicated by a circle.
In Regime III, we observe a string-like flow in which particles mostly move along lanes following the direction of shear. The system is then strongly anisotropic. The 
corresponding snapshot is shown in (d) for $T=0.003$ and $\dot \gamma = 4\times10^{-1}$ and an inset illustrates representative particle trajectories in the bulk of 
the system over a strain change $\Delta\gamma= 1.2$.}
\label{fig:phasediagram}
\end{figure*}

The phase diagram of the simulated model in the non-equilibrium steady state is summarized in Fig.~\ref{fig:phasediagram}(a). 

As the temperature $T$ and the 
shear rate $\dot \gamma$ are varied, the system can be found in three different regimes. 
{\it Regime I:} At small $\dot \gamma >0$ and small $T$, we observe a plastic flow with the  nucleation of free dislocations. Crystalline positional quasi-long-range order 
is then broken but hexatic quasi-long-range order persists. This is a flowing hexatic phase. A representative snapshot is shown in Fig.~\ref{fig:phasediagram}(b). 
Theories of $2d$ crystals under shear~\cite{bruinsma1982motion,ladd1983plastic} can be applied in this regime, especially in the limit of infinitesimal 
$\dot \gamma$ where they help discussing if and how a perfect crystal flows~\cite{sausset2010solids,nath2018existence,reddy2020nucleation}. 
{\it Regime II:} As $\dot \gamma$ or $T$ is increased, there is a transition to a regime where the dislocations are unbound and free disclinations are nucleated. Thus, 
both positional and bond-orientational correlations have a short-ranged spatial decay (see a snapshot in Fig.~\ref{fig:phasediagram}(c)). This regime is a flowing liquid which appears 
rather isotropic. 
{\it Regime III:} When $\dot \gamma$ is further increased, the imposed shear rate dominates the dynamics and we find a cross-over to a string-like flow, in which the 
particle motion mostly follow lanes in the direction of the imposed shear. This can be seen in the snapshot shown in Fig.~\ref{fig:phasediagram}(d) and in the associated inset, where some 
representative particle trajectories are displayed. In this regime the system is strongly anisotropic. 

In the subsequent sections, we provide a detailed characterization of the three regimes.

\section{Do two-dimensional crystals flow under an infinitesimal shear rate?}

\subsection{Theoretical arguments}

The fact that an infinitesimal shear stress destroys a solid phase by making it flow was theoretically established in full generality in Ref.~\cite{sausset2010solids}. The main idea is 
that a shear stress deforms a solid, thus inducing an extensive increase of the energy of the system. Such an excess energy can be relaxed at any finite temperature 
by {\it nucleating} droplets of the undeformed solid within the deformed solid state. Applying this metastability-nucleation argument one can conclude that an 
infinitesimal shear stress always destabilizes a solid state. The drawback of this treatment is that it provides a possible mechanism for flow but not necessarily the 
most efficient one. Sengupta, Sollich, and coworkers~\cite{nath2018existence,reddy2020nucleation} have recently built on this approach. They have used thermodynamic arguments and predicted the presence of a nearby first-order transition between two crystals with the same symmetry but different mechanical 
response to evaluate the effective stress at which a perfect crystal typically yields, {\it i.e.}, has its first plastic event, as a function of the shear rate. They have focused  
on the transient behavior in the limit $\dot\gamma\to 0$. Here, we are more interested in the steady-state regime and in the specific mechanisms at play in $2d$ crystalline solids.

In the case of a $2d$ crystal, the arguments can be made more explicit by pinpointing the underlying mechanism that gives rise to the instability of the solid state~\cite{bruinsma1982motion,ladd1983plastic}. 
The starting point is provided by the study of dislocations -- the defects destroying quasi-long-range  positional order -- in the presence of shear stress. 
We here focus on the physics along the glide direction (shear direction) which is a more dominant (faster) process than the physics along the climb direction (perpendicular to the shear direction).
In a $2d$ crystal without shear there are no free dislocations. The reason is that a pair formed by a dislocation and an anti-dislocation ({\it i.e.}, a dislocation of opposite Burgers vector) 
at a distance $r$ is subjected to an effective attraction through a potential $U_0(r)$ (without shear). This potential increases logarithmically at large $r$ as $U_0(r)=\frac{K a_0^2}{4\pi} { \ln (r/a_0)}$, 
where $a_0$ is the inter-particle distance (or lattice constant) and $K$ an effective elastic constant. In the presence of a shear stress $\sigma$, the pair of dislocations is submitted 
to an additional force in the glide direction so that the effective potential becomes:
\begin{equation}
U(r)=U_0(r)-a_0(r-a_0)\sigma .
\end{equation}
Even for a very small stress $\sigma$, the potential now favors unbinding of the dislocations as the linear term prevails on the 
logarithmic attraction: $U(r)$ diverges to minus infinity for $r\rightarrow \infty$. The competition between logarithmic attraction and linear repulsion leads to a finite energy 
barrier $\Delta U=U(r_c)-U(a_0)$ with $r_c=Ka_0/(4\pi \sigma)$,  thus making unbinding at nonzero temperature a thermally activated process. By computing the barrier and assuming an Arrhenius-type law one can obtain at leading order of the rate $R$
per unit time and unit area for the dissociation of a pair of dislocations and the ensuing formation of free dislocations~\cite{bruinsma1982motion,ladd1983plastic},
\begin{equation}
\label{nucleation_rate}
  R \sim \frac{D_{||}}{a_0^4}\left(\frac{\sigma a_0^2}{k_BT}\right)^{\frac{K a_0^2}{4 \pi k_BT}}e^{-2E_c/k_BT}\,,
\end{equation}
where $D_{||}$ is the diffusion constant in the glide direction and $E_c$ a microscopic energy scale. The important (and leading) term in this expression is associated 
to the power-law dependence in $\sigma$. 

Due to this mechanism, at any nonzero temperature and for an arbitrary small shear stress, a finite (albeit very small) density of free dislocations $\rho_{\rm disl}$ is produced, thus 
destroying the quasi-long-range positional order. 
The rate equation for $\rho_{\rm disl}$ is written by
\begin{equation}
\frac{\partial \rho_{\rm disl}}{\partial t} = R - \langle v \rangle r_c \rho_{\rm disl}^2,
\label{eq:rate_equation}
\end{equation}
where $\langle v \rangle$ is the mean velocity of free dislocations in the glide direction, driven by shear stress $\sigma$.
The second term in Eq.~(\ref{eq:rate_equation}) treats the recombination process approximatly~\cite{bruinsma1982motion}. At the steady-state, $\rho_{\rm disl}$ is obtained by 
\begin{equation}
    \rho_{\rm disl}=\sqrt{\frac{R}{\langle v \rangle r_c}}.
    \label{eq:steady_state}
\end{equation}
Free dislocations are expected to show a Brownian motion under an external force by shear, and hence, using the Einstein relation, $\langle v \rangle$ is given by
\begin{equation}
\langle v \rangle = a_0 \sigma D_{||}/(k_B T).
\label{eq:einstein}
\end{equation}
A moving dislocation also leads to deformation of the solid. The associated strain rate 
is proportional to the density of dislocations~\cite{dahm1989dynamics},
\begin{equation}
    \dot \gamma \sim \rho_{\rm disl} \langle v \rangle.
    \label{eq:strain_rate_relaxation}
\end{equation}
One combines Eqs.~(\ref{nucleation_rate},\ref{eq:steady_state},\ref{eq:einstein},\ref{eq:strain_rate_relaxation}) and arrives at a relation between the strain rate $\dot \gamma$ and shear stress $\sigma$,
\begin{equation}
\label{eqn:BHZ_flow}
  \dot\gamma \sim  D_{||} \left(\frac{\sigma a_0^2}{k_BT}\right)^{\frac{Ka_0^2}{8 \pi k_BT}+1}\,.
\end{equation}
The viscosity is defined as $\eta=\sigma/\dot \gamma$, and thus one finds
\begin{equation}
\label{eqn2:BHZ_flow}
  \eta \sim  \eta_0\left(\frac{\sigma a_0^2}{k_BT}\right)^{-\frac{Ka_0^2}{8 \pi k_B T}}\,,
\end{equation}
where $\eta_0$ is a constant with dimension of viscosity. The two expressions in Eqs.~(\ref{eqn:BHZ_flow},\ref{eqn2:BHZ_flow}) can be combined to give
\begin{equation}
\label{eqn3:BHZ_flow}
 \log \left(\frac \eta{\eta_0}\right) \sim -\frac 1{1+(8\pi k_BT)/(K a_0^2)} \log \dot\gamma + {\rm O}(1).
\end{equation}

These equations show that an infinitesimal shear stress indeed leads to plastic flow of a crystal and to a very large but finite viscosity. The behavior of 
the viscosity is however singular. It diverges when $\sigma \rightarrow 0$ or $\dot\gamma\to 0$, contrary to what happens for a liquid in which a finite value of the viscosity is 
reached when $\sigma \to 0$.

\subsection{Numerical results}

\begin{figure}
\centering
\includegraphics[width=\columnwidth]{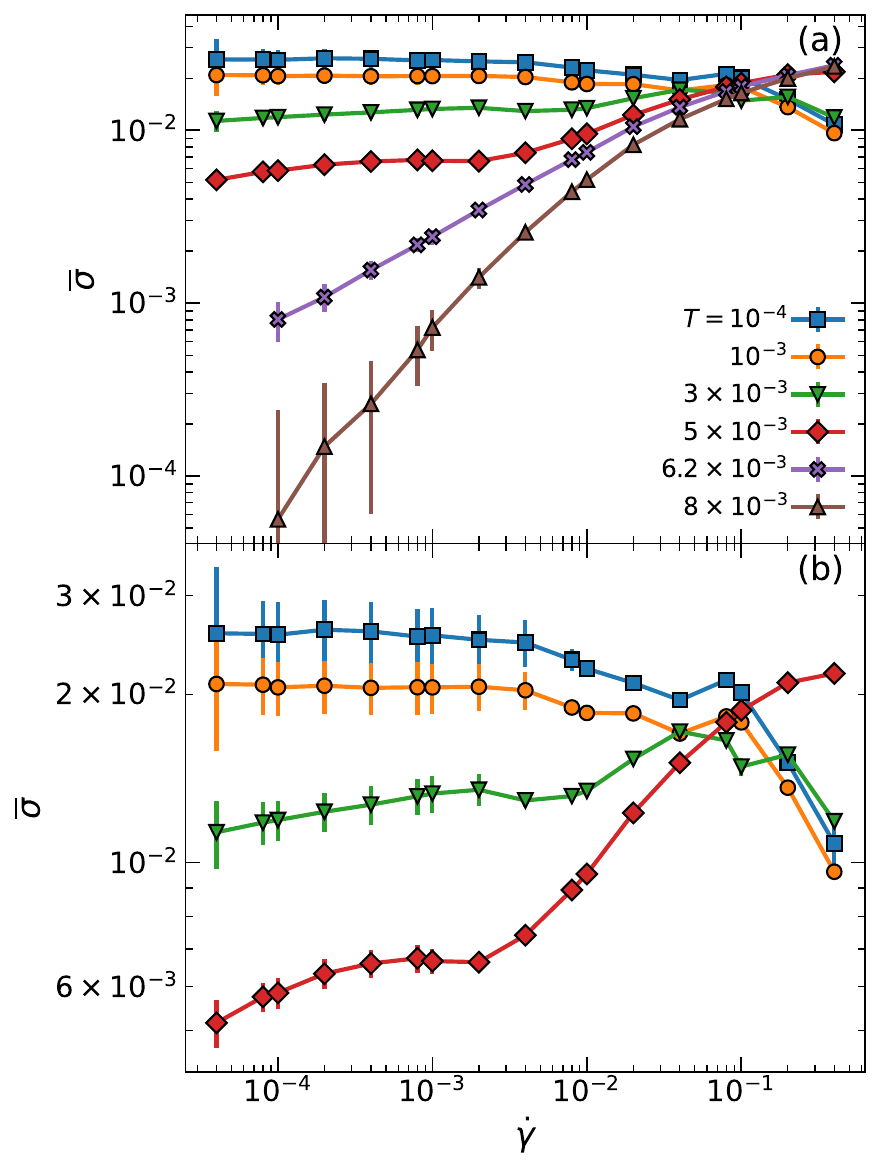}
\caption{Flow curves for a crystal of $N=14400$ particles under uniform simple shear. (a): Log-log plot of the averaged shear stress $\overline \sigma$ versus 
the shear rate $\dot\gamma$ for several temperatures. (b): Zoom-in plot of panel (a).}
\label{fig:flowcurves}
\end{figure}

We first measure the averaged shear stress $\overline \sigma$, where the overline denotes an average over time (or strain $\gamma$) and over independent 
trajectories in the steady state, as a function of the imposed shear rate $\dot \gamma$. The outcome is displayed on a log-log plot in Fig.~\ref{fig:flowcurves}(a) for 
more than three orders of magnitude of $\dot \gamma$ and a wide range of temperature from $T=0.0001$ to $0.0080$ that covers from the solid to the liquid 
phases found at $\dot \gamma=0$ (see above).

The flow curves at the lowest temperatures, $T=0.0001$ and $0.0010$, show a plateau at the smallest values of $\dot \gamma$ which indicates an apparent nonzero 
yield stress within our simulation time window. However, for the intermediate temperatures, $T=0.0030$ and $0.0050$, which are still below the estimated $T_{m,{\rm sol}}$ and 
thus correspond to a solid phase when $\dot \gamma = 0$, one clearly observes a steady decay of $\overline \sigma$ with decreasing $\dot \gamma$, as better  
seen in the zoomed-in plot of Fig.~\ref{fig:flowcurves}(b). Below some crossover shear-rate value, this decay is roughly linear on the log-log plot with a slope 
that decreases as $T$ decreases. This is compatible with the theoretical prediction in Eq.~(\ref{eqn:BHZ_flow}), which implies that 
$\log \overline \sigma \sim [1+Ka_0^2/(8\pi k_B T)]^{-1} \log \dot\gamma$ (but the data is not good enough to provide a meaningful extraction of the parameters), and 
supports the absence of a nonzero yield stress in the limit $\dot\gamma \to 0$. As $T$ is increased further,
$\overline\sigma$ decreases rapidly with decreasing $\dot \gamma$: one then enters the Newtonian fluid regime with no yield stress, as shown for instance in Fig.~\ref{fig:flowcurves}(a) for $T=0.0080$.
\begin{figure}
\centering
\includegraphics[width=\columnwidth]{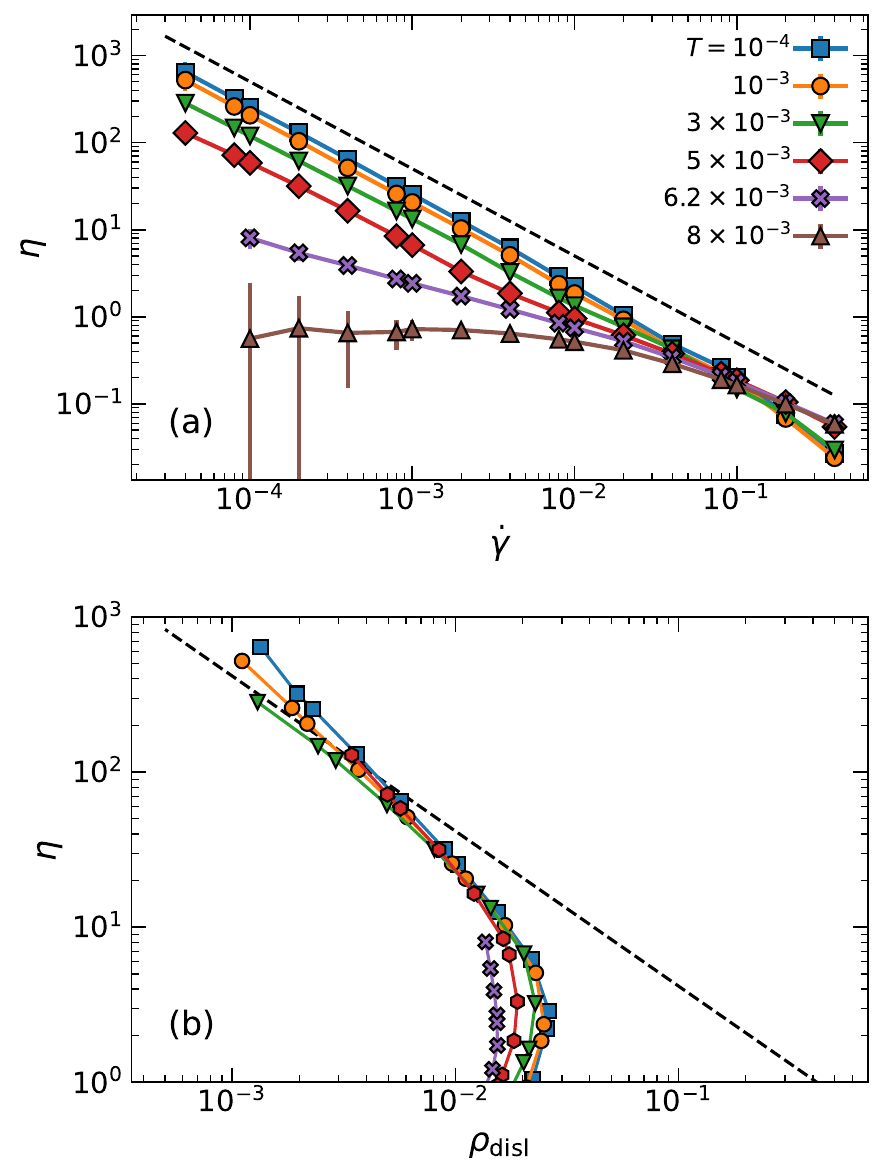}
\caption{
(a): Log-log plot of the effective viscosity  $\eta=\overline \sigma/\dot \gamma$ as a function of the shear rate $\dot\gamma$ for the same data as in 
Fig.~\ref{fig:flowcurves}(a). The dashed straight line shows the dependence $\eta \sim \dot \gamma^{-1}$.
(b): Log-log plot of the effective viscosity as a function of the density of dislocations $\rho_{\rm disl}$. The dashed line corresponds to $\eta \sim \rho_{\rm disl}^{-1}$.}
\label{fig:viscosity}
\end{figure}

To obtain a complementary picture we also plot the effective viscosity $\eta = \overline \sigma/ \dot \gamma$ in Fig.~\ref{fig:viscosity}(a). At low and intermediate 
temperatures, $T=0.0001-0.0050$, the data is well described by a power-law divergence at small $\dot \gamma$, $\eta \sim \dot \gamma^{-\alpha}$. As a 
consequence of the behavior of $\overline\sigma$ just described, we find that $\alpha = 1$ for the two lowest temperatures because of the apparent 
nonzero plateau found in $\overline\sigma$ within the simulation range, but it slightly deviates from $1$ for the two intermediate temperatures in agreement with 
a vanishing yield stress, and as expected from eq.(\ref{eqn3:BHZ_flow}). At the highest temperatures ($T=0.0080$), $\eta$ saturates toward a finite value, as expected for a Newtonian fluid. (At high shear rates 
the system displays shear thinning with a viscosity that decreases with increasing $\dot\gamma$ at all temperatures.) All the above results are illustrated for 
$N=14400$ but they weakly depend on system size: see Appendix~\ref{app:system-size}.
We also confirmed the absence of the yield stress and divergence of the viscosity in the SLLOD dynamics (see Appendix~\ref{app:SLLOD}).

According to the theoretical arguments recalled in the previous subsection, the plastic flow of a $2d$ crystal is driven by the nucleation of free dislocations 
induced by the stress (or the shear rate) and corresponding to the unbinding of dislocation/anti-dislocation pairs. The motion of the free dislocations relaxes 
the shear stress and it is more specifically predicted that the effective viscosity is inversely proportional to the density of free dislocations, 
$\eta = \sigma/ \dot \gamma \sim \rho_{\rm disl}^{-1}$ by using Eqs.~(\ref{eq:strain_rate_relaxation},\ref{eq:einstein}).
This is what leads to Eqs.~(\ref{eqn2:BHZ_flow},\ref{eqn3:BHZ_flow}). To more directly test the relation between the viscosity $\eta$ 
and the density of free dislocations $\rho_{\rm disl}$, we have determined the latter numerically, as explained in Appendix~\ref{app:disloc-disclin}. We show in 
Fig.~\ref{fig:viscosity}(b) a log-log plot of $\eta$ as a function of $\rho_{\rm disl}$. We find that data at different temperatures roughly collapse, and, although not perfect, a behavior compatible with $\eta \sim \rho_{\rm disl}^{-1}$ at high $\eta$ (or low $\dot \gamma$) 
is observed. This provides evidence that the mechanism for the divergence of the viscosity when $\dot\gamma \to 0$ is indeed the rarefaction of nucleated free 
dislocations. At lower $\eta$ or higher $\dot \gamma$, the data show a nonmonotonic dependence and the theoretical arguments no longer apply, as expected.

\section{Regime I: Flowing hexatic phase}

\subsection{Evidence for a hexatic phase and a shear-induced transition to a liquid phase}
\label{sub:hexatic}

We have seen that the crystaline solid at $\dot\gamma \to 0$ yields and flows as soon as an infinitesimal shear rate is imposed due to the nucleation of free dislocations. 
These free dislocations also disrupt the positional quasi-long-range order. Shear-induced melting of the crystal 
therefore take place as soon as $\dot\gamma \neq 0$. The question that remains is whether the flowing phase is a liquid with exponentially decaying 
translational and bond-orientational spatial correlations or an intermediate hexatic phase retaining quasi-long-range bond-orientational order.

\begin{figure}
\includegraphics[width=\columnwidth]{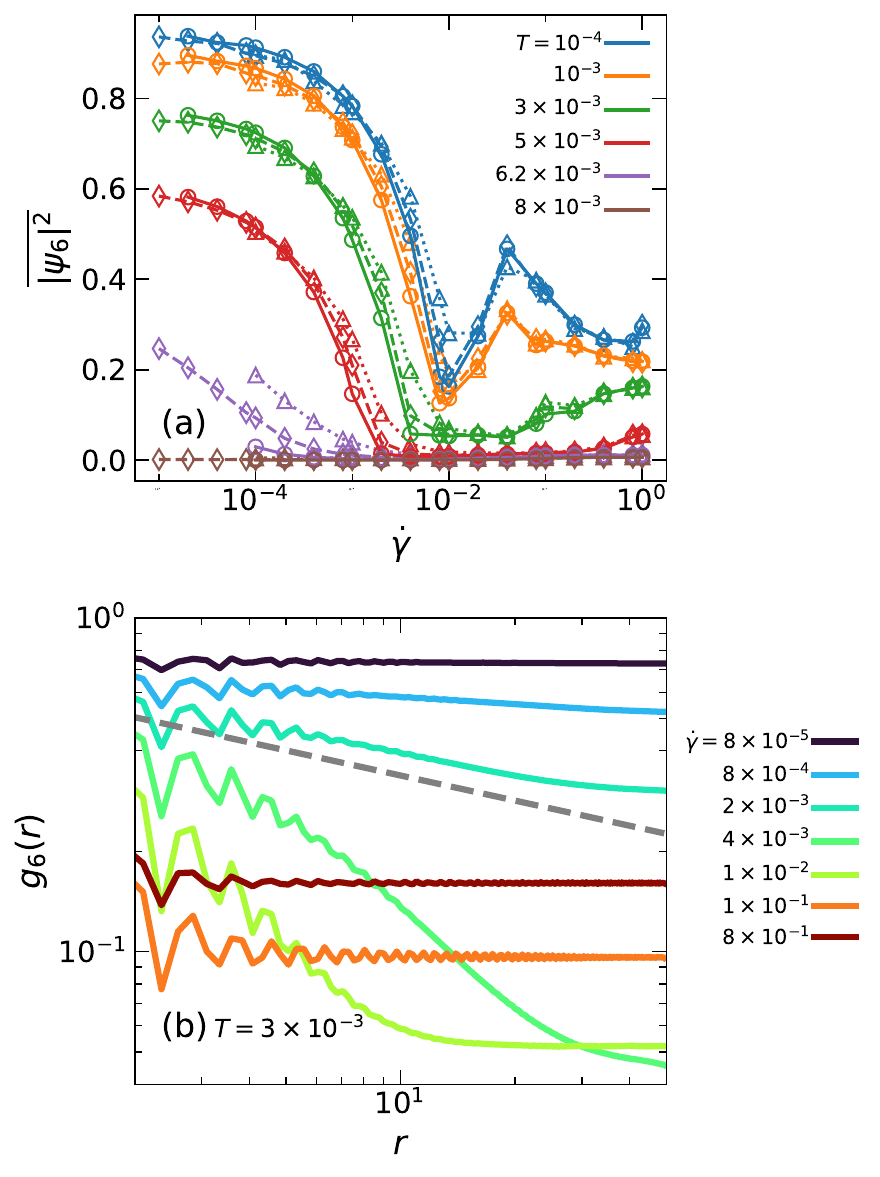}
\caption{(a): Averaged square modulus of the bond-orientational order parameter,  $\overline{|\psi_6|^2}$, as a function of shear rate for various temperatures and 
system sizes. Triangles (with dotted-line), diamonds (dashed-line), and circles (solid-line) correspond to data for $N=900$, $3600$, and $14400$, respectively. 
(b): Spatial decay of the bond-orientational correlation function $g_6(r)$ for $T=0.0030$, $N=14400$, and a wide range of $\dot \gamma$. The grey dashed line 
represents the bound imposed on a power-law decay by the KTHNY theory, $g_6(r) \sim r^{-1/4}$.}
\label{fig:psi6}
\end{figure}

We characterize the structural properties of the flowing phase by using the local $6$-fold bond-orientational local order parameter, 
\begin{equation}
\phi_{6,j} = \frac{1}{n_j}\sum_{k=1}^{n_j}e^{6i\theta_{jk}},
\end{equation}
where the sum is over the $n_j$ neighbors of particle $j$ that are determined through a Voronoi tessellation and $\theta_{jk}$ is the angle between the vector 
joining particle $j$ with particle $k$ and the (arbitrarily chosen) $x$-axis. From $\phi_{6,j}$ we compute the volume-averaged bond-orientational order 
parameter  $\psi_6=(1/N)\sum_{j=1}^N \phi_{6,j}$ and the $6$-fold bond-orientational spatial correlation function $g_6(r)$: see Appendix~\ref{app:BOorder} for 
more details.

We display in Fig.~\ref{fig:psi6}(a) the averaged square modulus of the  bond-orientational order parameter $\overline{|\psi_6|^2}$ versus $\dot \gamma$ for various temperatures and system sizes. For all temperatures in the solid and hexatic phases for the quiescent system ($\dot \gamma=0$), {\it i.e.}, for $T<T_{\rm m,hex}\approx 0.0062$, one finds that 
$\overline{|\psi_6|^2}$ decreases, first slowly and then in a quite rapid manner, as the shear rate increases and reaches a minimum before rising up again. However, 
one has to be careful about finite-size effects. Except for below $T_{\rm m,sol}$ with $\dot\gamma=0$ one indeed expects that $\overline{|\psi_6|^2}=0$ in the 
thermodynamic limit when the solid flows and free dislocations appear. As in the equilibrium hexatic phase, we expect that only quasi-long-range bond orientational order can 
be present. One then anticipates a dependence on the linear system size of the form $\overline{|\psi_6|^2}\sim L^{-\eta_6}$. Assuming that this flowing hexatic phase shares the same properties of its equilibrium counterpart one would then expect $\eta_6$ to be a temperature 
dependent anomalous dimension such that $\eta_6\leq 0.25$~\cite{nelson_review}. (Here, we make no difference between $L_x$ and $L_y$ because we have chosen them proportional to 
each other.) On the other hand in an isotropic liquid phase with only short-range order, $\overline{|\psi_6|^2}$ should decrease much more rapidly with system 
size, possibly as $L^{-1}$ because  the boundaries break the isotropy of space.

We indeed observe that at the smallest $\dot\gamma$, below some value that appears to decrease as temperature increases (but still stays below $T_{m, {\rm sol}}$), 
very little change of $\overline{|\psi_6|^2}$ takes place for the system sizes under study whereas at and around the minimum of $\overline{|\psi_6|^2}$ a visible decrease is found.
As shown in Fig.~\ref{fig:psi6min_vs_N} of Appendix \ref{app:system-size}, the minimum,  $\min_{\dot \gamma} \{  \overline{\lvert \psi_6\rvert^2} \}$, always decreases {more rapidly than $L^{-1/4}$ (and more so as $T$ increases because the system sizes as probably too small to reach the asymptotic regime at the lowest temperatures)}.
For $T=0.0062$, which is around $T_{m, {\rm hex}}$, the finite-size effects is strong even at low 
$\dot\gamma$ and for the highest temperature that always corresponds to a liquid phase $\overline{|\psi_6|^2}$ is always zero, at least up to a shear rate 
$\dot\gamma\sim 10^{-1}-10^0$. The data therefore indicate that a transition from a flowing hexatic phase to a liquid phase occurs at a shear rate that decreases 
as the temperature increases: This is the transition line between regimes I and II shown in Fig.~\ref{fig:phasediagram}(a).

The above results are also confirmed by looking at the bond-orientational correlation function $g_6(r)$. In Fig.~\ref{fig:psi6}(b), we illustrate the outcome for $T=0.003$ and a wide range of 
shear rates, but the results for all temperatures are given in Appendix~\ref{app:BOorder}. For the lowest rates $g_6(r)$ decays very slowly, as a power law 
$g_6(r)\sim r^{-\eta_6}$. The slope of the power law increases with $\dot\gamma$ and reaches the upper bound predicted by the KTHNY theory of the hexatic phase, 
{\it i.e.}, $\eta_6=0.25$, for some value slightly above $2\times10^3$. This suggests that the non-equilibirum transition at which the hexatic order is lost is in the same universality class of its equilibrium counterpart. 
For larger values, above $\dot\gamma=4\times10^3$, $g_6(r)$ decays quickly with 
an exponential rather than a power-law form. The passage from a power-law decay to an exponential decay is characteristic of a transition from quasi-long-range 
order to no order. This locates the transition between regimes I and II. Note that when $\dot\gamma$ increases further, typically above $10^{-1}$, $g_6(r)$ reaches 
a nonzero plateau at large distances suggesting the appearance of long-range bond-orientational order, but this will be discussed in the next section concerning 
regime III.

The disappearance of quasi-long-range bond-orientational order is due to the unbinding of dislocations and to the resulting appearance of free disclinations. This 
can be tested by identifying and characterizing the latter: see Appendix~\ref{app:disloc-disclin}. In Fig.~\ref{fig:disclinations}, we report for various temperatures and 
values of the shear rate the probability $p_{\rm disc}$ of finding at least one disclination in the sample during the plastic flow. It is zero when the system is in 
Regime I, which corresponds to a flowing hexatic phase with no free disclinations. At a rather well defined $\dot\gamma$ the probability jumps to a value of 1 
(or nearly 1 for the lowest temperatures) and the system is now in a (flowing) liquid phase. The onset of the jump corresponds to the boundary between regimes 
I and II shown in Fig.~\ref{fig:phasediagram}(a).

\begin{figure}
\includegraphics[width=\columnwidth]{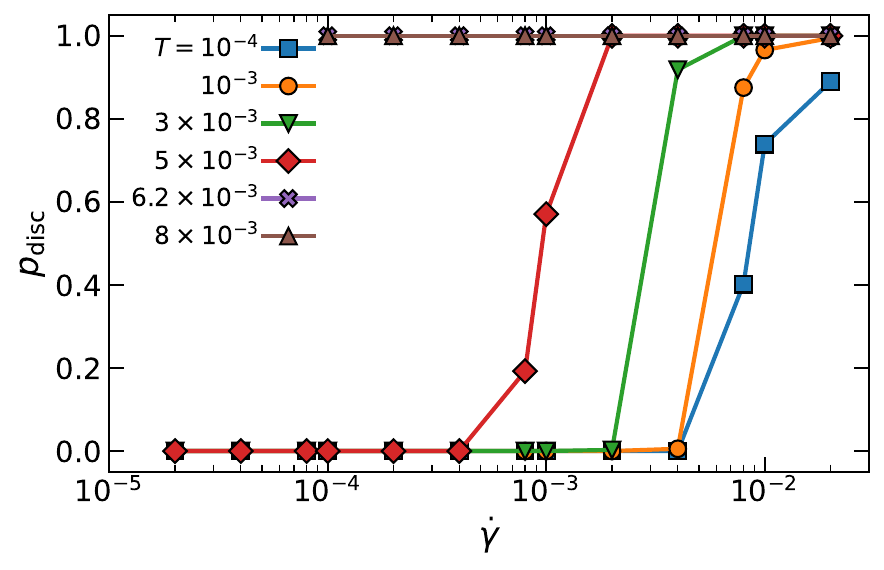}
\caption{Probability to find at least one free disclination in the system, $p_{\text{disc}}$, as a function of $\dot \gamma$ and $T$ for $N=14400$ particles.}
\label{fig:disclinations}
\end{figure}

By studying the $6$-fold bond-orientational order and the emergence of free disclinations (which are defects in this order) we have identified a transition between 
Regime I, which can be described as a flowing hexatic phase, and Regime II, which corresponds to a flowing liquid phase. This is in line with the findings of previous 
numerical simulations~\cite{weider1993shear,delhommelle2004simulations} and experiments~\cite{stancik2004dynamic,ramos2004shear} on $2d$ sheared 
crystals. However, we are not able to determine if the transition is continuous or first-order-like (as argued by Ref.~\cite{delhommelle2004simulations}).This aspect requires further investigations with huge comuputational efforts.

\subsection{Rotating crystals}
\label{sub:rotating}

In Regime I where quasi-long-range bond orientational order is present we have also studied the dynamics of the system in the steady state at fixed shear rate 
$\dot\gamma$. We have monitored the evolution with strain $\gamma$ (which parametrizes time) of several quantities. As previously observed in a simulation~\cite{weider1993shear} and an experimental~\cite{stancik2004dynamic} study of a sheared $2d$ crystal, we find evidence for a coherent rotation 
of hexagonal crystalline domains. Their size scales like the system size and, as argued above and further below, the phenomenon should therefore 
be taken as a finite-size effect that would likely not persist in this form in the thermodynamic limit.

\begin{figure}
\centering
\includegraphics[width=\columnwidth]{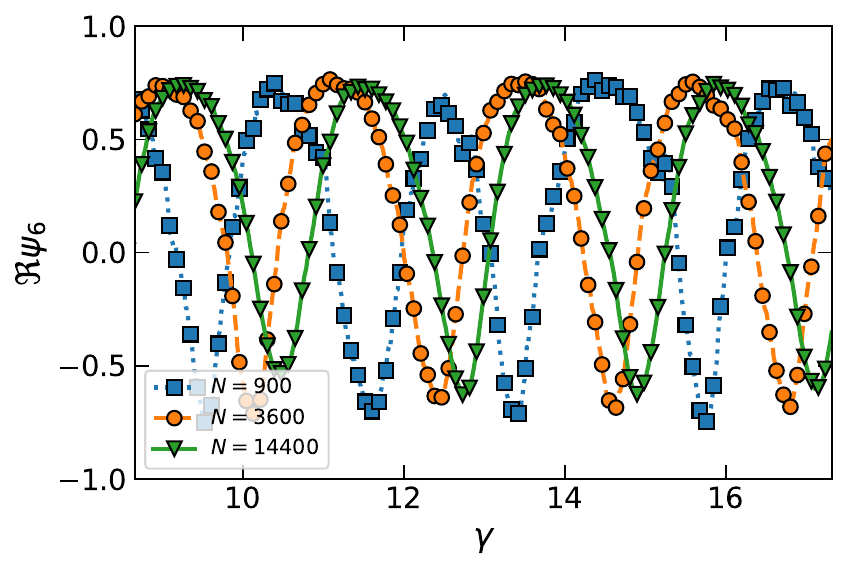}
\caption{Real part of the bond-orientational order parameter $\Re\{\psi_6\}$ obtained in a single trajectory as a function of the shear strain $\gamma$ for a fixed shear 
rate $\dot\gamma=1\times10^{-3}$ and temperature $T=0.0030$ (corresponding to Regime I). Different system sizes $N$ are shown.}
\label{fig:psi6oscillation}
\end{figure}

We first consider the (instantaneous, {\it i.e.}, not time averaged) $6$-fold bond-orientational order parameter, whose real part $\Re\{\psi_6\}$ as a function of $\gamma$, as shown in Fig.~\ref{fig:psi6oscillation}. One can see a clear oscillating behavior between a positive maximum value and a negative minimum one. The period  $\gamma^*$ of 
the oscillations can be estimated from a simple argument. Consider a hexagonal lattice that coherently rotates in a periodic box when the box is sheared at a rate 
$\dot\gamma$. The corresponding bond-orientational order parameter $\psi_6$ then periodically oscillates with a period $\tau^*$ which is such that 
$\tau^* \dot\gamma/2= \pi/3$. As by definition $\gamma^*=\dot\gamma\tau^*$, this immediately gives
\begin{equation}
    \gamma^* = \frac{2\pi}{3} \approx 2,
\end{equation}
which indeed captures well the oscillation period shown in Fig.~\ref{fig:psi6oscillation}.

\begin{figure*}
\centering
\includegraphics[width = 1.0\textwidth, keepaspectratio]{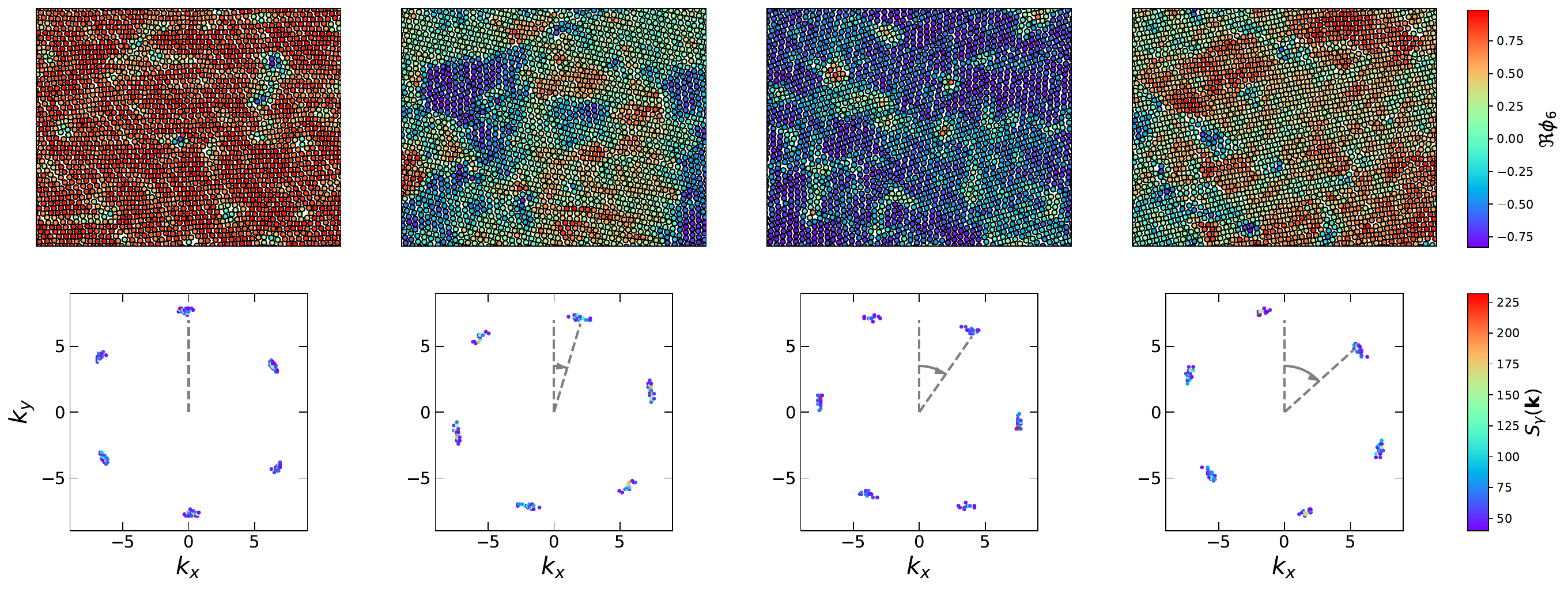} 
\caption{Crystal-like rotation as seen from real-space snapshots (top panels) and the associated instantaneous static structure factor $S_\gamma({\bf k})$ (bottom panels) for strain values $\gamma=9.0, 9.7, 10.1$, and $10.5$ (from left to right) which correspond to the maximum, the decreasing section, the minimum and the increasing section of the oscillation shown in Fig.~\ref{fig:psi6oscillation}. The snapshots are colored according to the value of the real part of the local bond-orientational order parameter, $\Re\{\phi_{6, j}\}$. The system size is $N=3600$, the temperature $T=0.0030$, and the shear rate $\dot\gamma=1 \times 10^{-3}$ 
(Regime I).}
\label{fig:rotation}
\end{figure*}

The rotation can also be directly seen by looking at the evolution of a given sample: real-space snapshots are displayed in the top panels of  Fig.~\ref{fig:rotation}. 
Particles are colored according to the value of the real part of the local bond-orientational order parameter $\phi_{6, j}$. When $\Re\{\phi_{6, j}\}=1$, the local environment of a particle is that of a perfect hexagonal triangular lattice with direction parallel to the $x$-axis, while when $\Re\{\phi_{6, j}\}=-1$, the orientation of the surrounding environment is rotated by an angle of $\pi/2$. The periodic appearance of red (large positive $\Re\{\phi_{6, j}\}$) and blue (large negative $\Re\{\phi_{6, j}\}$) regions indicates that 
the solid flows with a coherent rotation.

Another signature of coherently rotating crystalline domains is obtained by considering the instantaneous static structure factor $S_\gamma(\mathbf{k})$ measured from each snapshot~\cite{weider1993shear,stancik2004dynamic}. It is defined as
\begin{equation}
\label{eq:Sk}
S_\gamma(\mathbf{k}) = \frac{1}{N}\sum_{j,k=1}^N e^{i\mathbf{k}\cdot\left(\mathbf{r}_j - \mathbf{r}_k \right)},
\end{equation}
where $\mathbf{k} = (k_x, k_y) = (2\pi n_x/L_x, 2\pi n_y/L_y)$, with $n_x, n_y$ integers, consistently with the imposed periodic boundary condition. In the solid 
phase in thermal equilibrium, this function shows six peaks in the $(k_x, k_y)$ plane that are located on the vertices of a regular hexagon. In the bottom panels of 
Fig.~\ref{fig:rotation} one can see that the $6$-fold pattern rotates while the deformation proceeds, indicating that the local environment of each particle is 
coherently rotated during the flow. As already mentioned such a crystal rotation has been observed in two-dimensional colloid experiments~\cite{stancik2004dynamic} and a SLLOD molecular-dynamics simulation~\cite{weider1993shear}. It was also recently predicted as a consequence of dislocation nucleation in a mesoscopic athermal model~\cite{baggio2023inelastic}.  

Several comments are in order. First, the oscillations are not quite symmetric between the vicinity of the maxima of $\Re\{\psi_6\}$ and that of the minima (see 
Fig.~\ref{fig:psi6oscillation}). The rotation is faster and the absolute value is smaller near the minima, which corresponds to the situation where the crystal-like 
domains are oriented perpendicularly to the shear direction (see also the experimental result in Ref.~\cite{stancik2004dynamic}). Second, the overall coherence 
of crystal rotation does not mean that the particles themselves rotate coherently as they can escape the crystalline structure and be replaced by other ones. Finally, 
we recall once again that a rotating crystal, characterized by a nonzero bond-orientational order parameter, even an instantaneous one, is likely a finite-size effect. 

\begin{figure}
    \centering   \includegraphics[width=\columnwidth]{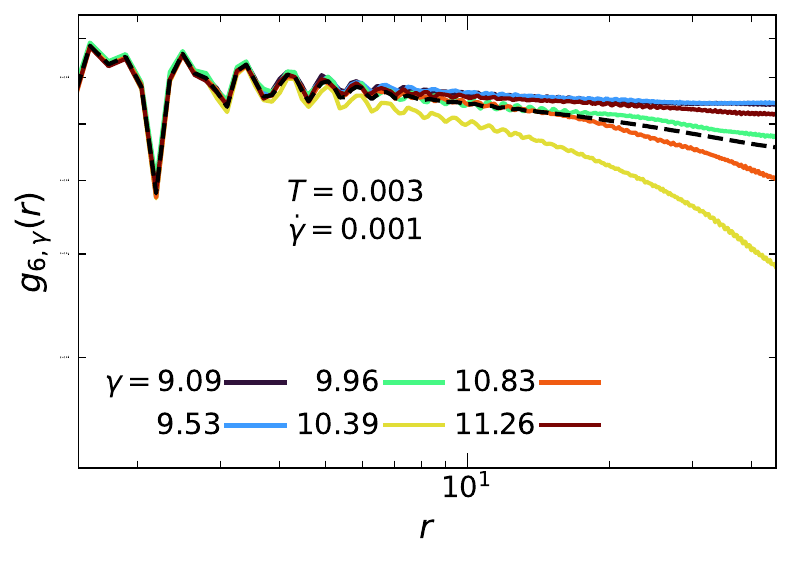}
    \caption{Instantaneous value of the bond-orientational correlation function $g_{6,\gamma}(r)$ for several  values of the strain $\gamma$ (solid colored lines) and its value averaged over a period (black dashed line) for a system of $N=14400$ particles at $T=0.0030$ and $\dot \gamma=10^{-3}$ (Regime I).}
    \label{fig:rotation_g6inst}
\end{figure}

Interestingly, we observe an oscillating behavior also in the instantaneous value of the bond-orientational correlation function, $g_{6,\gamma}(r)$, as shown in Fig.~\ref{fig:rotation_g6inst}. This correlation function passes from an increasingly steep power-law decay to an exponential one, coming back to the power-law decay at the end of one period. This suggests that the flow of the rotating solid proceeds through a transient melting of the sample. This is similar to what was found experimentally on sheared colloids~\cite{stancik2004dynamic}. The average value of the correlation function across one oscillation period nevertheless displays a power-law decay (see the dashed line in Fig. \ref{fig:rotation_g6inst}), suggesting that only quasi-long-range bond-orientational order is present in instantaneous configurations in the thermodynamic
limit.

\section{Crossover to string-like flow}
\label{sec:string-like}

\begin{figure}
\includegraphics[width = \columnwidth]{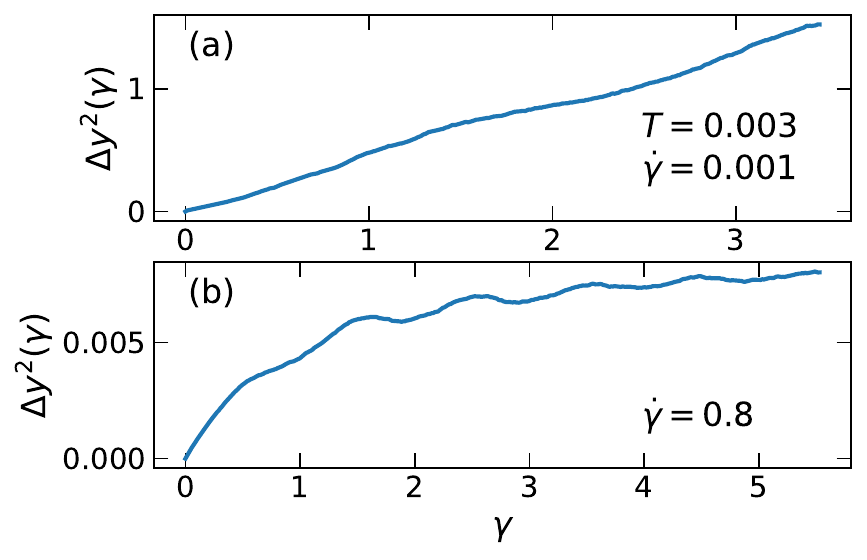}
\caption{Mean square displacement $\Delta y^2(\gamma)$ along the direction perpendicular to the shear for one trajectory in the steady state as a function of the strain $\gamma$ for two different 
shear rates,  $\dot\gamma=10^{-3}$ (a) and $\dot\gamma=0.8$ (b), at a temperature $T=0.0030$. $\gamma$ is measured from a configuration in the steady state. The top panel corresponds to Regime I and the bottom one to Regime III.}
\label{fig:transverseMSD}
\end{figure}

The isotropic flowing liquid phase (Regime II) appears rather narrow at low temperature and widens as $T$ is increased, as seen from Figs.~\ref{fig:phasediagram}(a) 
and \ref{fig:psi6}(a). Indeed, upon further increase of $\dot\gamma$, the imposed shear dominates the dynamics of the system and one finds a crossover to a situation 
in which particles in the steady state flow along bands parallel to the shear direction. This leads to a string-like flow (Regime III), as seen in the real-space snapshot 
of Fig.~\ref{fig:phasediagram}(d). The effect of an increased shear rate on the ability of particles to diffuse in the direction perpendicular to the shear is presented in 
Fig.~\ref{fig:transverseMSD}, where we plot the mean square displacement in the $y$ direction as a function of strain for a fixed temperature $T=0.0030$ and two different shear rates. While the mean square displacement grows linearly for the small shear rate (which corresponds to the flowing hexatic phase of Regime I) as expected for a diffusive motion, it is virtually constant for the large shear rate corresponding to the string-like 
flow of Regime III.

\begin{figure}
\includegraphics[width = \columnwidth]{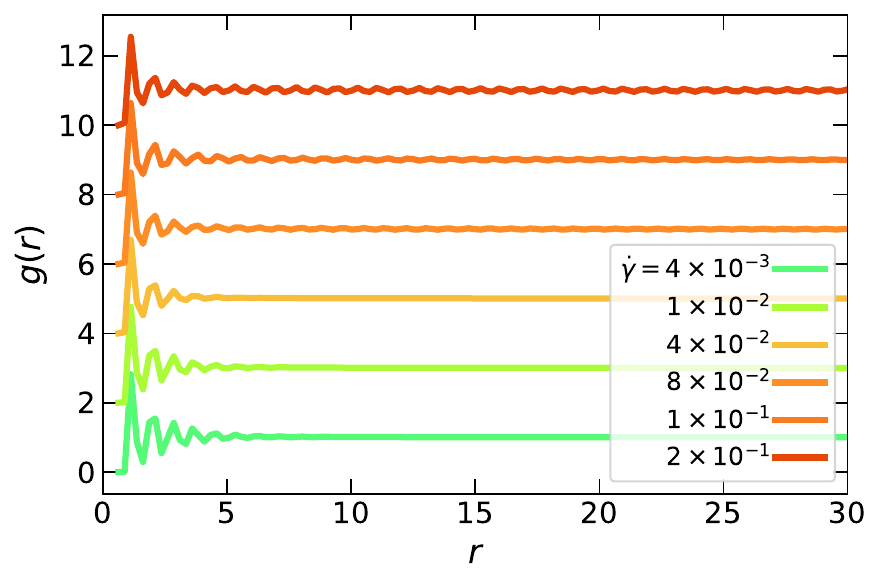}
\caption{Radial distribution function $g(r)$ for a system of $N=14400$ particles at a temperature $T=0.0030$ and for several shear rates $\dot \gamma$ covering 
the three regimes of flow. The data for different $\dot \gamma$ are shifted along the $y$-axis for clarity.}
\label{fig:g_T0.0030}
\end{figure}

Several signatures of the new regime are found in the structure. One can see from Fig.~\ref{fig:psi6}(a) that the averaged square modulus of the bond-orientational 
order parameter starts to increase again to nonzero values (with virtually no system-size dependence). Accordingly, the bond-orientational correlation function reaches 
a nonzero plateau at large distances: see Fig.~\ref{fig:psi6}(b). One can also look at the radial distribution function (averaged over all directions) $g(r)$. It is 
plotted for $T=0.0030$ for several $\dot \gamma$ covering all three regimes in Fig.~\ref{fig:g_T0.0030}. For the smallest $\dot \gamma$, $g(r)$ quickly decays to one, 
as expected from the lack of positional order in Regimes I and II. However, for $\dot \gamma \gtrsim 8\times 10^{-2}$, a series of ripples appear, which persist up to 
the system size. More data are presented in Appendix~\ref{app:radial}, which allows us to estimate the crossover line between regimes II and III as a function of 
temperature. The obtained phase boundary is shown in Fig.~\ref{fig:phasediagram}(a).

Note that the ripples in $g(r)$ do not imply positional order characteristic of a crystal. It instead signals that the flow is organized in parallel bands along the shear 
direction. Beyond the real-space snapshots, this is supported by the study of the transverse static structure factor that probes the ordering of the particles in the 
direction orthogonal to the flow. As illustrated in Fig.~\ref{fig:Sky} of Appendix~\ref{app:transverse}, this clearly shows an organization of the particles in bands of width roughly equal to the particle size, in agreement with the visualization provide by Fig.~\ref{fig:phasediagram}(d).

\begin{figure}
\includegraphics[width = \columnwidth]{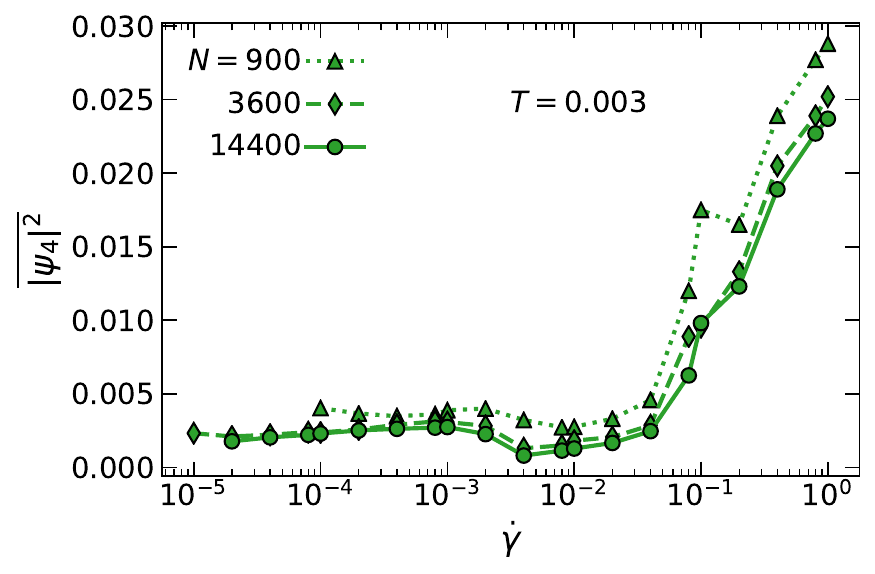}
\caption{Averaged square modulus of the $4$-fold bond-orientational order parameter,  $\overline{|\psi_4|^2}$, as a function of shear rate for a temperature 
$T=0.0030$ and several system sizes. Triangles, diamonds, and circles correspond to data for $N=900$, $3600$, and $14400$, respectively.}
\label{fig:psi4}
\end{figure}

The regime of string-like flow is highly anisotropic. This is what explains the nonzero value of the $6$-fold bond-orientational order parameter presented in Fig.~\ref{fig:psi6}(a). This is confirmed by 
the study of another bond-orientational order parameter, {\it e.g.}, that associated with cubic ($4$-fold) symmetry, 
\begin{equation}
\psi_4=\frac 1N\sum_{j=1}^N \frac{1}{n_j}\sum_{k=1}^{n_j}e^{4i\theta_{jk}}.
\end{equation}
We plot in Fig.~\ref{fig:psi4} the averaged square modulus of $\psi_4$ as a function of the shear rate $\dot\gamma$ for several system sizes and a temperature 
$T=0.0030$. One can clearly see that the flowing system ceases to be isotropic (even if there might be a shear-induced small distortion of the 
structure~\cite{westermeier2016connecting,vermant2005flow} possibly associated with the boundaries and leading to the small finite-size effect 
seen in the figure)  around $\dot\gamma\sim 10^{-1}$, which corresponds to the beginning of Regime III (see Fig.~\ref{fig:phasediagram}(a)).

The existence of a string-like regime of flow has also been reported in a $2d$ colloid experiment at higher shear rate~\cite{stancik2004dynamic}. On the 
other hand, it has not been found in molecular dynamics simulations up to rates for the order of $10^{-1}$~\cite{weider1993shear,delhommelle2004simulations}.
Inertial effects which are absent in colloidal systems and in our Brownian dynamics 
simulations therefore appear to suppress the string-like organization of the flow at high shear rate.

\section{Conclusion}

We have given a unified description of a two-dimensional crystal under a constant shear rate, starting from the detailed account of how a  perfect equilibrium solid yields and flows when an infinitesimal shear rate is imposed and then mapping out the whole phase diagram of the flowing steady state in 
the plane formed by temperature and shear rate. In doing so, we have carefully discussed the role of the topological defects (dislocations and disclinations) 
and of the finite-size effects.

Shear-induced melting of the $2d$ crystal proceeds in two steps: the  solid loses long-range bond-orientational order and flows for an infinitesimal shear rate 
(in the thermodynamic limit) and the resulting flowing hexatic phase then melts to a flowing (rather isotropic) liquid at a finite shear rate that depends on 
temperature. Finally, at high shear rate, a third regime corresponding to a strongly anisotropic string-like flowing phase appears. We note that contrary to what 
has been suggested~\cite{stancik2004dynamic} the phase diagram does not seem to be controlled by a single dimensionless parameter such as the 
P\'eclet number, which for Brownian dynamics is simply proportional to $\dot\gamma/T$. Indeed, one can see from Fig.~\ref{fig:phasediagram}(a) that a 
large $\dot\gamma$ and a small $T$ do not have the same effect so that for the same ratio the system can be found in any of the three regimes.

What remains to be done in two dimensions is a precise characterization of the nature of the transition from the flowing hexatic to the flowing liquid. This would 
require using much larger system sizes to check whether the transition is continuous or rather first-order-like with a coexistence between the two different 
flowing phases~\cite{engel2013hard,kapfer2015two}. In case of a continuous transition, it is important to determine whether the universality class is the same one of the equilibrium case. 
Beyond this, an obvious extension is to investigate yielding and shear melting of three-dimensional crystals (for a review, see Ref.~\cite{vermant2005flow}) which have been theoretically shown to flow at infinitesimal shear rate in the thermodynamic 
limit~\cite{sausset2010solids,nath2018existence,reddy2020nucleation} but for which no intermediate hexatic-like phase exists in equilibrium. 
Finally, it would be interesting to study how the flow properties of crystals identified in this paper change and converge to the rheology of amorphous materials~\cite{nicolas2018deformation} when introducing size polydispersity systematically~\cite{shiba2010plastic, kawasaki2011plastic} {or whether the connection made between the  mechanical properties of dense active matter and sheared amorphous solids~\cite{morse2021direct} carries over to crystalline phases.}

\begin{acknowledgements}
We thank J. Sethna for discussions. This work was supported by the Simons Foundation Grant No. 454935 (G.B.).
\end{acknowledgements}

\appendix

\section{Shear stress measurement}
\label{app:shear-stress}

We measure the $xy$ component of the stress tensor denoted as $\sigma$ by using the Irving-Kirkwood formula~\cite{irving1950statistical} for the overdamped Brownian 
Dynamics,
\begin{equation}
    \sigma = -\frac{1}{A}\sum_{i,j} x_{ij}\left( \frac{\partial v(\mathbf{r}_{ij})}{\partial \mathbf{r}_{ij}}\right)_y,
\end{equation}
where $A=L_xL_y$ is the area of the system, $x_{ij} = x_i-x_j$, with $x_i$ the position of particle $i$  along the $x$-axis (according to the minimum image convention), 
and $-(\partial v/\partial \mathbf r_{ij})_y$ is the $y$ component of the force exerted by particle $j$ onto particle $i$. Note that when evaluating the 
distance $\mathbf{r}_{ij}$ we take into account the periodic boundary condition and the minimum image convention. We recall that $x$ is the direction of the imposed 
shear.

When we use the SLLOD dynamics (see Appendix~\ref{app:SLLOD} for details), the shear stress $\sigma_{\rm SLLOD}$ contains an extra term due to momentum flow:
\begin{equation}
    \sigma_{\rm SLLOD} = \sigma + \frac{1}{A}\sum_{i} \frac{p_{x, i}p_{y, i}}{m} ,
\end{equation}
where $\mathbf{p}_i = (p_{x, i}, p_{y, i})$ is the momentum of particle $i$ (see Eq.~\eqref{eq:SLODD_conf}).

\section{Results from nonequilibrium SLLOD molecular dynamics simulations}
\label{app:SLLOD}

\begin{figure}
    \includegraphics[width=\columnwidth]{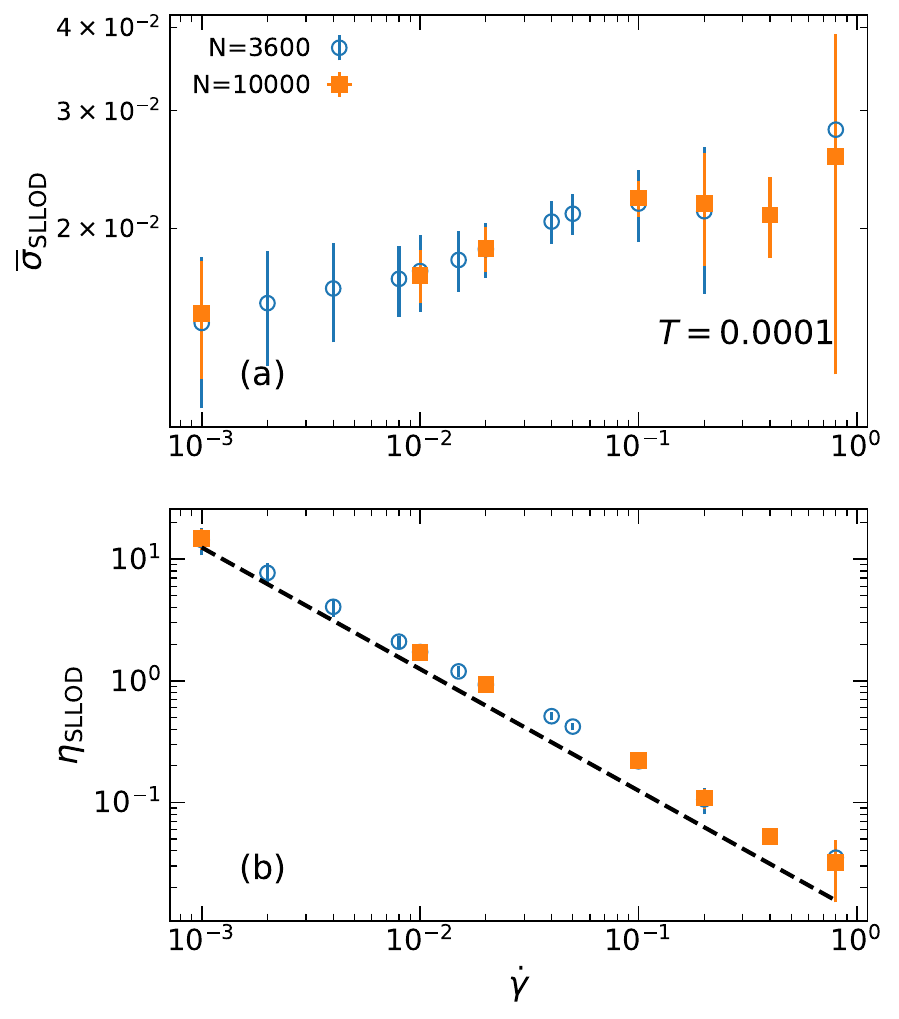}
    \caption{(a) Flow curve of the two-dimensional crystal undergoing the SLLOD dynamics at $T=0.0001$ for a wide range of the strain rate for $N=3600$ and $10000$. (b) Corresponding effective viscosity. The black dashed line represents the divergence of the viscosity as a power law, $\eta \sim \dot \gamma^{-1}$.}
    \label{fig:SLLOD_flowcurve}
\end{figure}

In order to confirm the genericness  of the conclusions in the main text, in particular, the absence of a  yield stress and the divergence of the effective viscosity when $\dot \gamma \to 0$, we have also used the SLLOD dynamics as an alternative to the Brownian dynamics. We follow the implementation developed in Ref.~\cite{costa2013transient}.

We first explain the implementation of the thermostat in the nonequilibrium simulations.
The imposed shear field leads the system to overheat and, therefore, a thermostat mechanism is needed. A general prescription for the development of a thermostat is as follows~\cite{frenkel2023understanding}: One defines a ``heat bath'' coordinate, say $\zeta$, which is coupled to the equations of motion. Such a dynamics must sample the system in a chosen state or ensemble. This condition determines the form of the coupling between the thermostat and the particles. The choice of the coupling is not unique. In particular, when the thermostat is applied out of equilibrium, some choices can introduce a bias toward certain regimes with respect to others (for a discussion relevant to the present problem, see Ref.~[\onlinecite{delhommelle2004simulations}]). In this paper, we use for simplicity a configurational thermostat~\cite{braga2005configurational}. The configurational temperature, labeled $T_{\rm conf}$, is measured from the configuration of the particles in real space and their interactions:  
\begin{equation}\label{TconF}
  k_B T_{\text{conf}} = \frac{\sum_{i}\left(\frac{\partial U}{\partial \mathbf{r}_i}\right)^2}{\sum_{i} \frac{\partial^2 U}{\partial \mathbf{r}_i^2}}, 
\end{equation}
where $U$ is the total potential energy of the system.
The equation of motion for the SLLOD dynamics coupled with the configurational thermostat are as follows~\cite{costa2013transient}:
\begin{equation}
    \begin{split}
        \label{eq:SLODD_conf}
        \dot {\bf r}_i & = \frac{\mathbf{p}_i}{m} + \dot\gamma \left(y_i- \frac{L_y}{2}\right) \mathbf{e}_x - \zeta \frac{\partial U}{\partial \mathbf{r}_i}  \\
        \dot {\bf p}_i & = -  \frac{\partial U}{\partial \mathbf{r}_i} - \dot\gamma p_{y, i} \mathbf{e}_x\\
        \dot \zeta & = \frac{F_\zeta}{M_\zeta} \\
        F_\zeta &= \sum_{i=1}^{N} \left( \frac{\partial U}{\partial \mathbf{r}_i} \right)^2 - k_BT\sum_{i=1}^{N} \frac{\partial^2 U}{\partial \mathbf{r}_i^2}
    \end{split}
\end{equation}
Here, $\zeta$ is the coordinate of the thermostat, $F_\zeta$ the force governing its evolution, and $M_\zeta$ its ``mass''. A velocity Verlet-like integration scheme \cite{costa2013transient} has been implemented:
\begin{equation}
    \begin{split}
        \mathbf{r}_i(t+\Delta t) =& \mathbf{r}_i(t) + \Delta t \left(\frac{\mathbf{p}_i(t)}{m} + \dot\gamma\left(y_i-\frac{L_y}{2}\right)\mathbf{e}_x\right) \nonumber \\
        & + \Delta t\left(\zeta(t) + \frac{\Delta t}{2m}\right)\mathbf{F}_i(t)  \\
        \mathbf{p}_i(t+\Delta t)  =& \mathbf{p}_i(t) + \frac{\Delta t}{2}\left(\mathbf{F}_i(t) + \mathbf{F}_i(t + \Delta t)\right) \\
        & + \frac{\Delta t \dot\gamma}{2}\left(p_{y,i}(t+\Delta t) + p_{y,i}(t)\right)\mathbf{e}_x \\
        \zeta(t + \Delta t) =& \zeta(t) + \frac{\Delta t}{2M_{\zeta}}\left( F_\zeta(t) + F_\zeta(t+\Delta t)\right),
    \end{split}
\end{equation}
where $\mathbf{F}_i = -\frac{\partial U}{\partial \mathbf{r}_i}$ is the force acting on particle $i$ due to the interaction with the other particles. Time is measured in units of $\tau_0 = \sqrt{\frac{md^2}{\epsilon}}$.
We report results obtained through the SLLOD dynamics for systems of $N=3600$ and $10000$ particles at $T=0.0001$. Using a time step $\Delta t=0.01$ and a thermostat mass $M_\zeta=0.1$. We have chosen the units of mass $m$ such that $\tau_0=1$. 

Figure~\ref{fig:SLLOD_flowcurve}(a) shows the flow curves, $\overline \sigma$ as a function of $\dot\gamma$. We see no evidence of a yield stress as the average stress appear to keep decreasing at the lowest shear rates. The decrease of $\overline{\sigma}$ with $\dot\gamma$ is enhanced by the presence of inertia with respect to Brownian Dynamics.
The corresponding viscosity plot is shown in Fig.~\ref{fig:SLLOD_flowcurve}(b). We see a power-law divergence of $\eta$ approaching $\dot \gamma \to 0$.
These results are consistent with those obtained with the Brownian dynamics and presented in the main text.

\section{System size dependence}
\label{app:system-size}
In this Appendix, we report results on the different system sizes investigated by the Brownian dynamics. 

\begin{figure}
\includegraphics[width=\columnwidth]{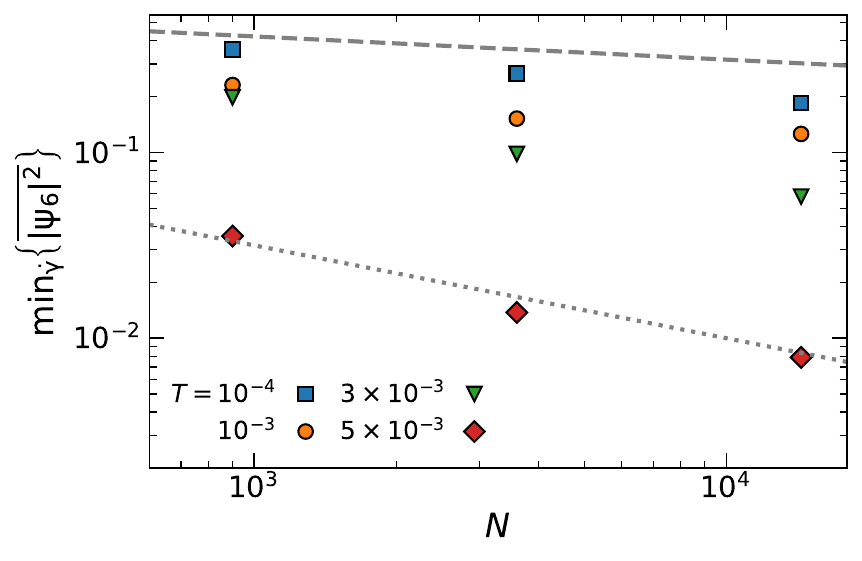}
\caption{System-size dependence of $\min_{\dot \gamma} \{\overline{\lvert \psi_6\rvert^2}\}$, the minimum value over $\dot\gamma$ reached by $\overline{\lvert \psi_6\rvert^2}$ in Fig.~\ref{fig:psi6}(a), for various temperatures below the putative $T_{m, {\rm sol}}$. The dashed and dotted lines indicates a $L^{-1/4}$ and a $L^{-1}$ dependence, respectively.}
\label{fig:psi6min_vs_N}
\end{figure}

Figure~\ref{fig:psi6min_vs_N} displays the variation with the system size $N$ of the minimum over $\dot\gamma$ of $\overline{\lvert\psi_6\rvert^2}$ (shown in Fig.~\ref{fig:psi6}(a) of the main text) for several temperatures. As discussed in the main text, the decrease with $N$, shown here on a log-log plot, is always more rapid than $L^{-1/4}$, which is the limiting behavior for a hexatic phase. One can observe that the slope associated with the apparent power law is steeper as the temperature increases.  

\begin{figure}
    \includegraphics[width=\columnwidth]{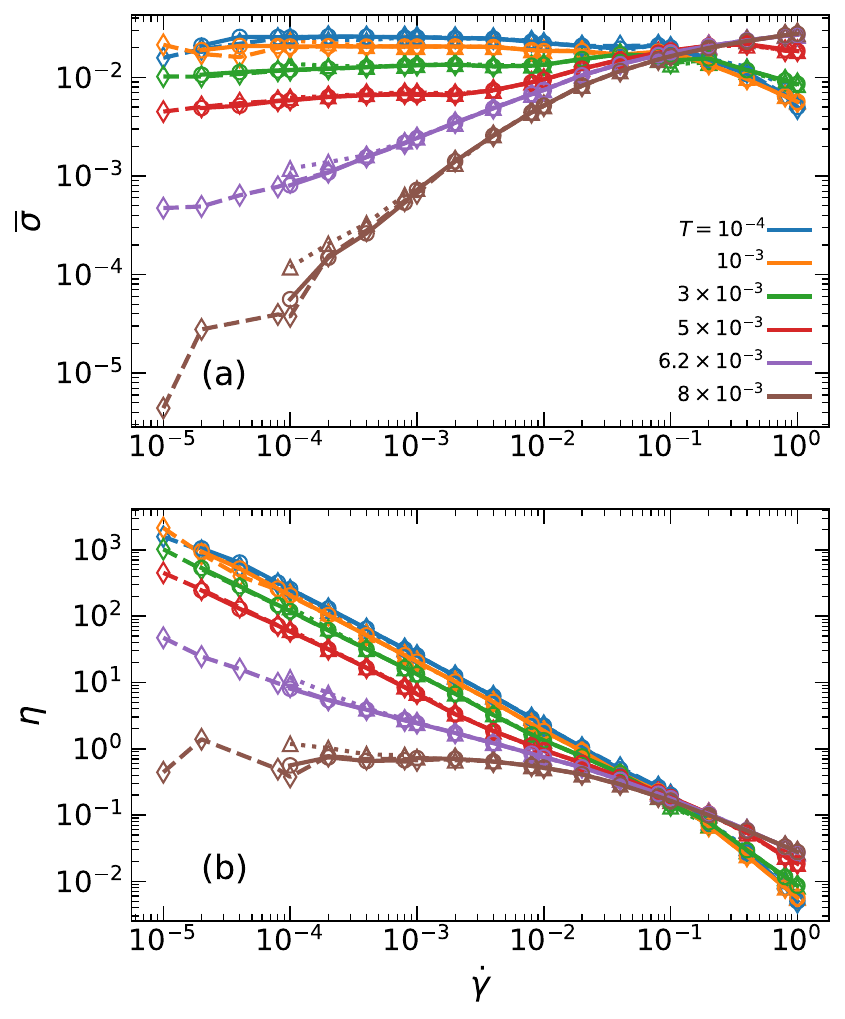}
    \caption{Flow curves obtained from the Brownian dynamics for the averaged shear stress $\overline \sigma$ (a) and the effective viscosity $\eta$ (b)  for several system sizes $N$. Triangles (with dotted-line), diamonds (dahsed-line), and circles (solid-line) correspond to data for $N=900$, $3600$, and $14400$, respectively.}
    \label{fig:size_effect_flowcurve}
\end{figure}

We also plot the flow curves and the corresponding viscosity for different system sizes, $N=900$, $3600$, and $14400$, in Fig.~\ref{fig:size_effect_flowcurve}.
We do not find any significant system-size dependence in these quantities.

\section{Identification of dislocations and disclinations}
\label{app:disloc-disclin}

Disclinations and dislocations are point topological defects in two dimensions: disclinations are defects in the bond-orientational order and dislocations in the positional 
order.

The starting point to identify disclinations is to perform a Voronoi tessalation of the given configuration of particles (snapshot). From the construction we count the number 
of neighbors of each particle. At low temperatures most particles have 6 neighbors (the average number of neighbors is constrained to be 6 in $2d$ Euclidean space) and some 
have 5 or 7 neighbors. Particles with a number of neighbors different than 6 correspond to disclination defects. The defect organization is illustrated in 
 Fig.~\ref{fig:phasediagram}(b-d) of the main text.  We have checked that the concentration of disclinations corresponding to particles with more than 7 neighbors and 
 less than 5 neighbors are negligible in the conditions that we study.

Dislocations are dipoles formed by two disclinations of opposite topological charge. They can be identified with a pair of adjacent 5-fold and 7-fold coordinated particles. 
In practice, however, dislocations can be condensed, forming clusters, {\it e.g.}, grain boundaries, and 5- and 7-fold particles can also appear close to each other 
at vacancies~\cite{digregorio2022unified}. In order to detect truly isolated dislocations and disclinations, we introduce a cutoff radius $r_{\rm cut}$. If no 5- or 7-fold coordinated particle is found 
within a distance $r_{\rm cut}$ from a putative dislocation (respectively, disclination), this dislocation (resp., disclination) is considered as isolated or free. The 
cutoff distance $r_{\rm cut}$ is  separately chosen for dislocations and disclinations, as described below. 

For the identification of free disclinations, a natural cutoff $r_{\rm cut}$ is the first minimum of the radial distribution function (see below for its definition), which can be 
taken as a characterizing the notion of adjacency for two particles. We thus set $r_{\rm cut}=1.5$. We have checked that the results do not change significantly when 
varying $r_{\rm cut}$ from $1.0$ to $2.0$. In Fig.~\ref{fig:disclinations} of the main text, we show the probability $p_{\rm disc}$ of finding at least one free disclination in a given configuration. At lower and 
intermediate temperatures ($T=0.0001-0.0050$) and low $\dot \gamma$, $p_{\rm disc}$ is zero since all disclinations are bound in dislocations, while $p_{\rm disc}$ 
very rapidly increase at some larger $\dot \gamma$ to reach a value close to 1. We limit the display of data to $\dot\gamma\leq 2 \times 10^{-2}$ since,
at higher shear rates,  the concentration of defects is large and the identification of the isolated disclinations becomes meaningless. 

\begin{figure}
\centering
\includegraphics[width = \columnwidth]{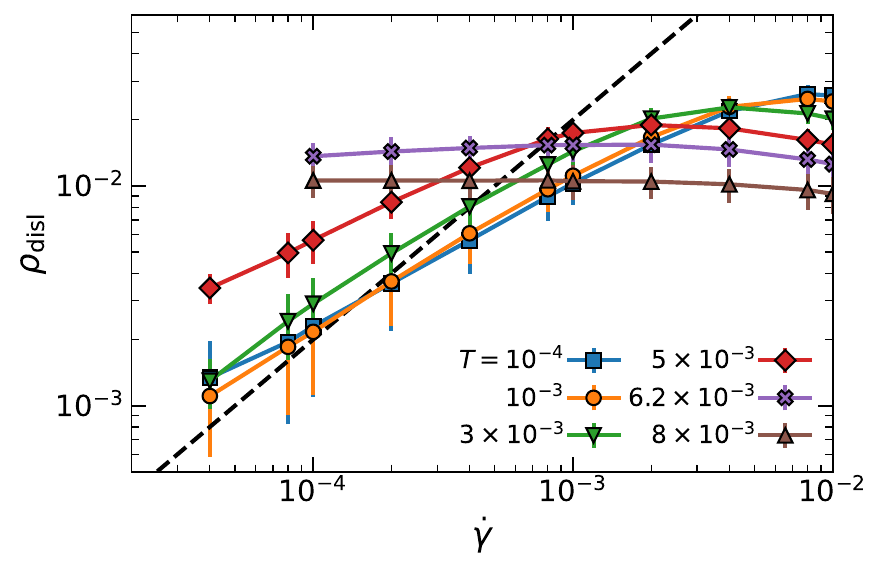}
\caption{Density of free dislocations, $\rho_{\rm disl}$, as a function of the shear rate $\dot\gamma$ for $N=14400$.
The dashed straight line corresponds to $\rho_{\rm disl} \sim \dot \gamma$.}
\label{fig:dislocations}
\end{figure}

For defining free dislocations, we choose a cutoff distance $r_{\rm cut}=2.5$, close to the second minimum of the radial distribution function ({\it i.e.}, beyond the 
second coordination shell around a given particle). Figure~\ref{fig:dislocations} shows the resulting density of free dislocations, $\rho_{\rm disl}$, for various values of 
$\dot \gamma$ and $T$. At lower and intermediate temperatures ($T=0.0001-0.0050$), $\rho_{\rm disl}$ roughly linearly increases with $\dot \gamma$ for low 
$\dot \gamma$, as argued in Eq.~(\ref{eq:strain_rate_relaxation})~\cite{dahm1989dynamics}. We limit the display of data to $\dot \gamma \leq 10^{-2}$ because for higher $\dot \gamma$, the concentration of the defects is so large that 
identifying isolated dislocations becomes difficult and meaningless. As $T$ is increased, $\rho_{\rm disl}$ increases, and the dependence on the shear rate saturates.
The measured $\rho_{\rm disl}$ is used in Fig.~\ref{fig:viscosity}(b) of the main text. We have also varied $r_{\rm cut}$ from 1.0 to 2.5 and confirmed that $\rho_{\rm disl}$ 
is insensitive to $r_{\rm cut}$ in Regime I, thereby showing that the relation between the viscosity and the density of free dislocations in Fig.~\ref{fig:viscosity}(b) is robust.

\begin{figure}
    \includegraphics[width=\columnwidth]{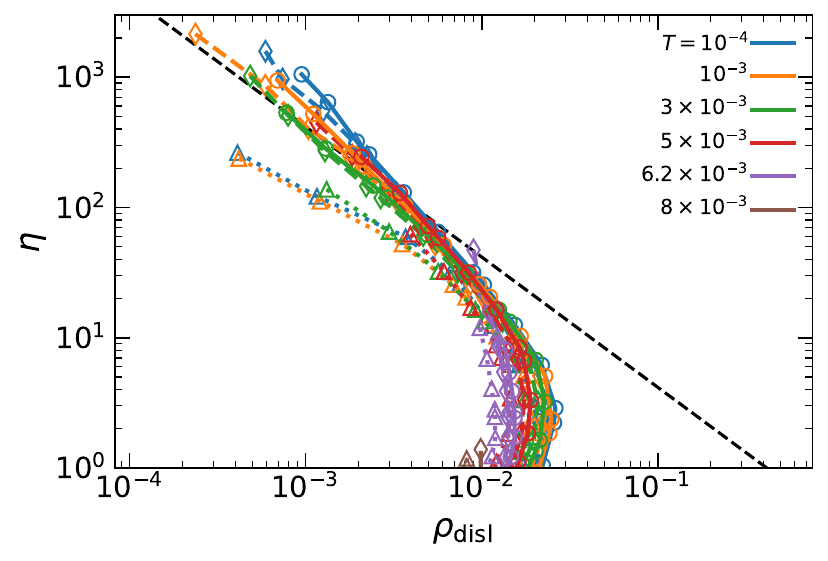}
    \caption{Viscosity $\eta$ of the system as a function of the dislocation density $\rho_{\rm disl}$ for various system sizes. Triangles (with dotted-line), diamonds (dashed-line), and circles (solid-line) correspond to data for $N=900$, $3600$, and $14400$, respectively.}    
    \label{fig:size_effect_dislocations}
\end{figure}

We also report the system size dependence of the viscosity $\eta$ versus dislocation density $\rho_{\rm disl}$ curve in 
Fig.~\ref{fig:size_effect_dislocations}.
We see that finite size effects suppress the dislocation density at $N=900$. Yet, these effects do not appear when comparing data for $N=3600$ and $N=14400$, consolidating our conclusions in the main text.

\section{Radial distribution function}
\label{app:radial}

\begin{figure*}
    \centering
   \includegraphics[width =2.\columnwidth]{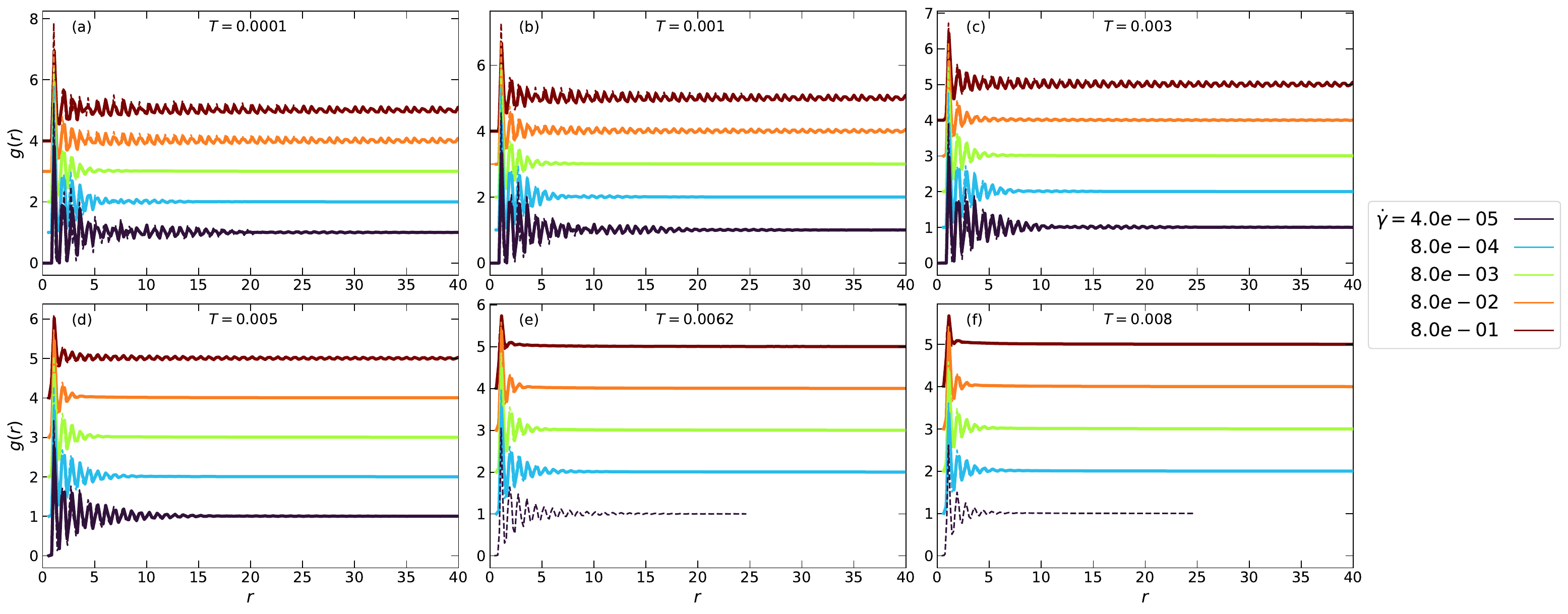}
    \caption{Radial distribution function $g(r)$ for systems with $N=3600$ (dashed curve) and $N=14400$ (solid curve) particles for various values of 
    $T$ and $\dot \gamma$. $g(r)$'s are shifted vertically by hand for clarity. }
    \label{fig:g}
\end{figure*}

The radial distribution function, $g(r)$, is computed according to
\begin{equation}
g(r) = \frac{A}{2 \pi r \Delta r N(N-1)} \sum_{i,j, (i\neq j)}^N \overline{
\int_r^{r+\Delta r}\delta(r'-|{\bf r}_{ij}|)dr'} ,
\end{equation}
where $\delta(x)$ is the Dirac delta function, $A=L_xL_y$ is the area of the system, and $\Delta r$ is the width of the bin used in the numerical evaluation. We 
take $\Delta r\approx 0.16$ for $N=3600$, $\Delta r\approx0.25$ for $N=14400$, and $\Delta r\approx 0.28$ for $N=57600$. The overline denotes the average over time and trajectories in the steady state.

In Fig.~\ref{fig:g} we show $g(r)$ for all the temperatures investigated and some representative values of the shear rate $\dot\gamma$. The onset $\dot\gamma$ 
corresponding to the appearance of system-spanning ripples is used for the phase boundary between Regime II and III in Fig.~\ref{fig:phasediagram}(a).

\section{Bond-orientational order parameter and its spatial correlations}
\label{app:BOorder}

\begin{figure*}
\centering
\includegraphics[width=2.\columnwidth, keepaspectratio]{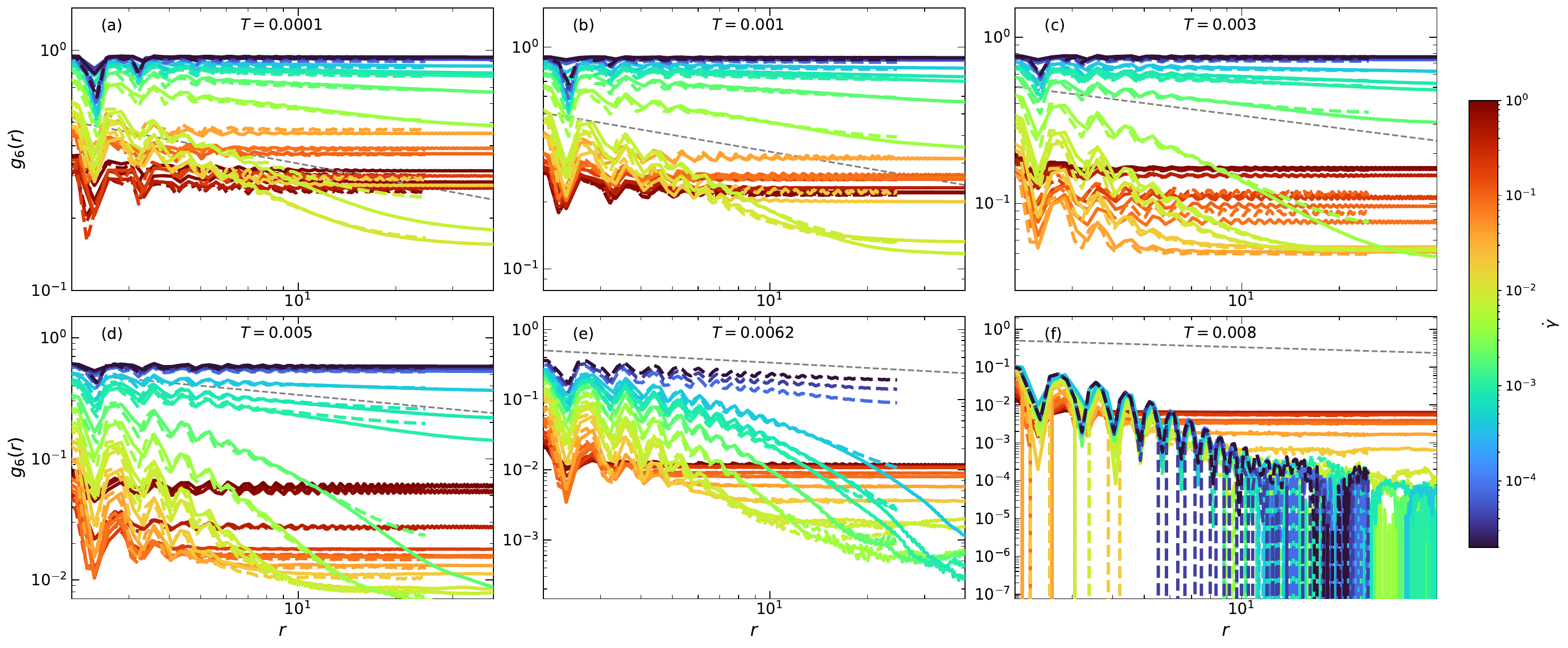}
\caption{$6$-fold bond-orientational correlation function, $g_6(r)$, for a system of $N=3600$ (dashed curves) and $N=14400$ (solid curves) particles. The gray 
dashed straight lines in the background represent the upper bound imposed on the exponent $\eta_6$ of the power-law decay for a hexatic phase by the KTHNY 
theory, $g_6(r) \sim r^{-1/4}$. }
\label{fig:g6}
\end{figure*}

We study the local $6$-fold bond-orientational order parameter for each particle $j$,
\begin{equation}
    \phi_{6,j} = \frac{1}{n_j}\sum_{k=1}^{n_j}e^{6i\theta_{jk}},
\end{equation}
where the sum is over the $n_j$ neighbors of particle $j$ that are determined through a Voronoi tessellation and $\theta_{jk}$ is the angle characterizing the 
vector (the ``bond'') joining particles $j$ and $k$, which is determined through the relation $\cos \theta_{jk} = \mathbf{\hat r}_{jk} \cdot \mathbf{e}_x$, with 
$\mathbf{\hat r}_{jk} = \frac{\mathbf{r}_k - \mathbf{r}_j}{\lvert \mathbf{r}_k - \mathbf{r}_j \rvert }$ a vector of unit norm joining particle $j$ with particle $k$ and 
the $x$-axis is arbitrarily chosen.

From this local order parameter, one can define the volume-averaged bond-orientational order parameter,
\begin{equation}
    \psi_6 = \frac{1}{N}\sum_{j=1}^N \phi_{6,j}. 
\end{equation}
When the system has a perfect hexagonal structure, $|\psi_6| = 1$, while in a disordered liquid, $|\psi_6|$ is 
nearly zero.
We also define the $6$-fold bond-orientational spatial correlation function,
\begin{equation}
\begin{split}
g_6(r) &= \frac{A}{2 \pi r \Delta r N(N-1)g(r)} \\
&\times \sum_{i,j, (i\neq j)}^N \overline{
\int_r^{r+\Delta r} \phi_{6, i} \phi_{6, j}^* \delta(r'-|{\bf r}_{ij}|)dr'} ,
\end{split}
\end{equation}
where $\Delta r$ is defined as in the previous section and the correlation function is conventionally normalized by the radial (isotropic) distribution function $g(r)$ to 
remove some of the effects coming from local positional ordering. $\overline{\cdots}$ denotes an average over time (or strain) and independent trajectories 
once the steady state has been reached. 

In Fig. \ref{fig:g6}, we show the log-log plots of $g_6(r)$ for two system sizes and all values of $T$ and $\dot \gamma$ considered in this study.
At low temperatures, below the melting temperature $T_{m, {\rm sol}} \approx 0.0055-0.0060$, and small shear rates, $g_6(r)$ has a power law decay, $g_6(r) \sim r^{-\eta_6}$ 
with $\eta_6 \leq 0.25$, establishing the presence of hexatic quasi-long-range order (Regime I). For higher values of $\dot\gamma$, $g_6(r)$ decays faster than the 
KTHNY bound (Regime II). Upon raising $\dot\gamma$ even further but still at low temperatures, $g_6(r)$ displays small plateau, with some ripples, signaling a new 
flow regime. Figure~\ref{fig:g6} also shows the absence of significant finite-size effects as the curves for the two system sizes essentially coincide, except for the 
lowest values of $\dot \gamma$: then, the power-law decay of $g_6(r)$ seems to saturate for the smaller system size; this effect 
disappears when the system size increases, suggesting that it is a finite-size effect.

\begin{figure}[b]
    \includegraphics[width=\columnwidth]{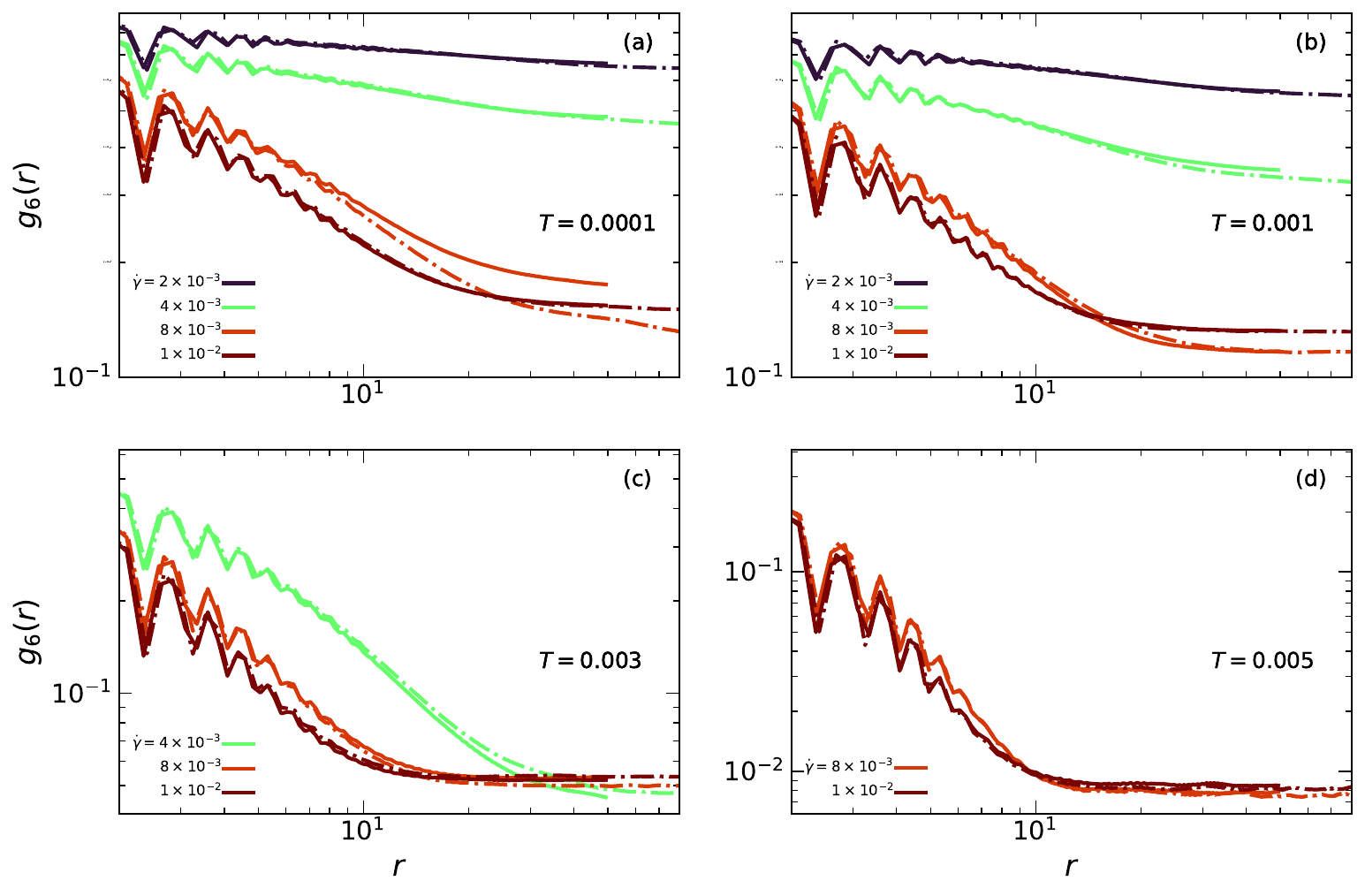}
    \caption{Orientational correlation function $g_6(r)$ for a system size of $N=14400$ (solid-lines) and $N=57600$ (dash-dotted-lines) in the vincinity of the transition between Regime I and II for several temperatures. }
    \label{fig:size_effect_g6}
\end{figure}

Additionally, we have performed simulations for a larger system of $N=57600$ particles in the vicinity of the Regime I-II transition to see the orientational correlation function $g_6(r)$ at a longer distance. The resulting plots are compared with the ones obtained for $N=14400$ particles in Fig.~\ref{fig:size_effect_g6}.
The results show little deviation between the two system sizes, except the trend that the smaller systems reach the plateau earlier at the hexatic quasi-long-range order regime (Regime I), as expected in generic spatial correlation functions. We note that the final plateau is also observed in the liquid regime without showing the system size dependence. This observation suggests that the plateau in the liquid regime is a genuine consequence of the anisotropy of the system, even in the thermodynamic limit.

\section{Transverse structure factor and string-like regime}
\label{app:transverse}

In this Appendix, we present more supporting evidence for the description of Regime III as a string-like flow in which particle motion is organized in parallel bands. 
\begin{figure*}
    \centering
    \includegraphics[width=2.0\columnwidth]{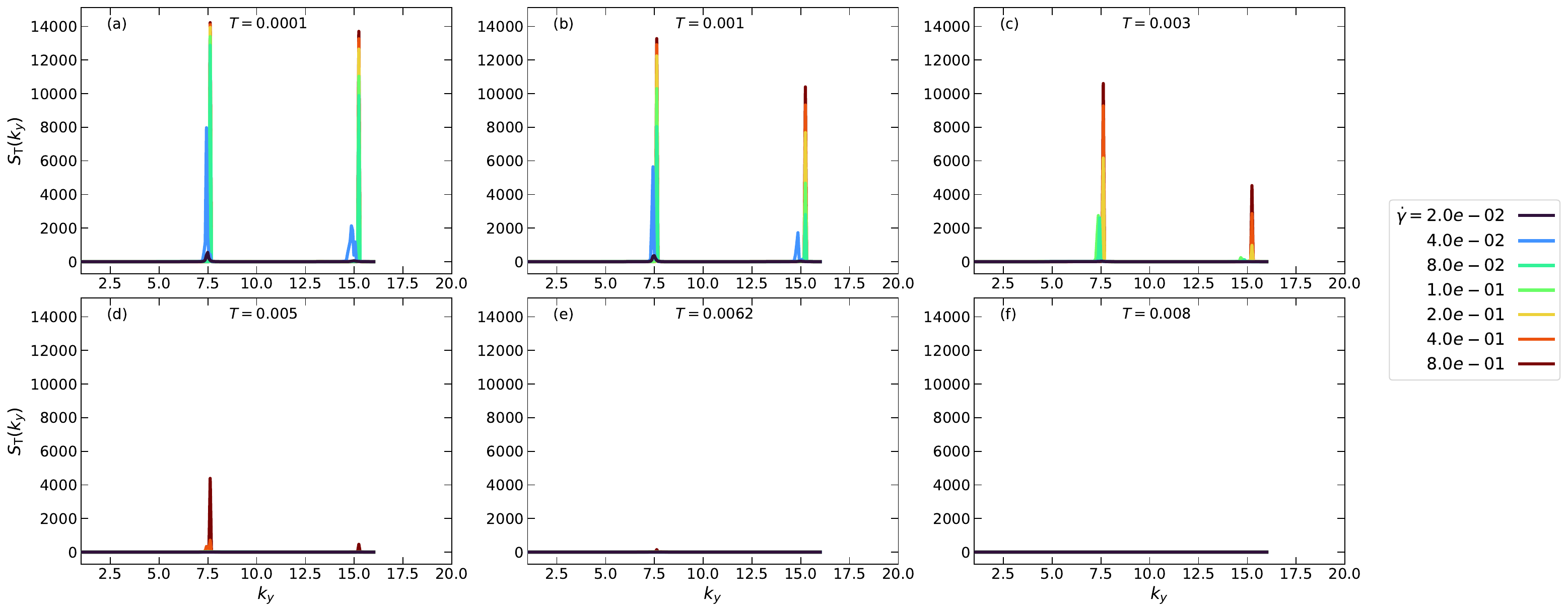}
    \caption{Structure factor for the direction transverse to the flow, $S_{\rm T}(k_y)$, for a system of $N=14400$ particles and various temperatures and shear rates. 
    As $\dot\gamma$ increases at low enough temperature, sharp  primary and secondary peaks appear near $k_y\approx 7.7$ and $k_y\approx 15.4$.}
    \label{fig:Sky}
\end{figure*}

We show in Fig. \ref{fig:Sky} the transverse structure factor computed for modes perpendicular to the direction $x$ of the shear flow,
\begin{equation}
    S_{\rm T}(k_y) = \frac{1}{N} \overline{ \sum_{j,k=1}^N e^{ik_y\left(y_j - y_k\right)}} ,
\end{equation}
with $k_y= 2\pi n_y/L_y$, $n_y$ being an integer.

As the shear rate increase (at low enough temperature), $S_{\rm T}(k_y)$ develops sharp primary and secondary peaks whose magnitude grows until it becomes 
of order $N$. This signals the appearance of string-like ordering induced by the flow (see the snapshot in Fig.~\ref{fig:phasediagram}(d)). The position of the first and second peak correspond respectively to 
$\frac{2\pi}{c_{0,y}}$ and $\frac{4\pi}{c_{0,y}}$, with $c_{0,y}$ the distance along the $y$ direction between the centers of the particles located in two adjacent rows on 
a triangular lattice. $c_{0,y}$ is related to the lattice constant $c_{0}$ by the relation $c_{0,y}=\frac{\sqrt{3}}{2}c_0$. 

\newpage
\bibliography{bibFede}

\end{document}